\documentclass[aps,prd,twocolumn,showpacs,amsmath,amssymb]{revtex4}
\usepackage[dvips]{graphicx}
\newcommand{\beq}{\begin{equation}}
\newcommand{\eeq}{\end{equation}}
\newcommand{\beqar}{\begin{eqnarray}}
\newcommand{\eeqar}{\end{eqnarray}}
\newcommand{\eps}{\varepsilon}
\newcommand{\barb}{\bar{b}}
\newcommand{\Fint}[1]{{\cal D}[#1]}
\newcommand{\VEV}[1]{\langle{#1}\rangle}
\newcommand{\chibar}{\bar{\chi}}
\newcommand{\mq}{{m_q}}
\newcommand{\Eq}{{E_q}}
\newcommand{\asig}{a_\sigma}
\newcommand{\bsig}{b_\sigma}
\newcommand{\gsig}{g_\sigma}
\newcommand{\sigM}{\sigma}
\newcommand{\VM}{\widetilde{V}_M}
\newcommand{\VB}{\widetilde{V}_B}
\newcommand{\FEFF}{{\cal F}_{{\rm eff}}}
\newcommand{\Feff}{F_{{\rm eff}}}
\newcommand{\FeffB}{F_{{\rm eff}}^{(b)}}
\newcommand{\FeffQ}{F_{{\rm eff}}^{(q)}}
\newcommand{\rhoB}{\rho_{\scriptscriptstyle B}}
\newcommand{\arcsinh}{\mathrm{arcsinh}\,}
\newcommand{\arccosh}{\mathrm{arccosh}\,}
\newcommand{\comment}[1]{}
\newcommand{\Psfig}[3]{\includegraphics[width=#1 #3]{PS/#2}}
\begin{document}
\title{
Phase diagram at finite temperature and quark density
\\
in the strong coupling limit of lattice QCD for color SU(3)
}
\author{N. Kawamoto, K. Miura, A. Ohnishi, T. Ohnuma}
\affiliation{
Division of Physics, Graduate School of Science, Hokkaido~University,
Sapporo 060-0810, Japan
}

\begin{abstract}
We study the phase diagram of quark matter
at finite temperature ($T$) and chemical potential ($\mu$)
in the strong coupling limit of lattice QCD for color SU(3).
We derive an analytical expression of the effective free energy
as a function of $T$ and $\mu$, including baryon effects.
The finite temperature effects are evaluated
by integrating over the temporal link variable exactly in the Polyakov gauge
with anti-periodic boundary condition for fermions.
The obtained phase diagram shows 
the first and the second order phase transition at low and high
temperatures, respectively, and those are
separated by the tri-critical point in the chiral limit.
Baryon has effects to reduce the effective free energy
and to extend the hadron phase to a larger $\mu$ direction
at low temperatures.
\end{abstract}


\maketitle

\section{Introduction}
\label{Sec:Introduction}

Exploring various phases of quark and nuclear matter
has recently attracted much attention
both theoretically and experimentally.
In the Relativistic Heavy-Ion Collider (RHIC) experiments,
it is probable that
strongly interacting Quark Gluon Plasma
is created in heavy-ion collisions~\cite{QM2004}.
The phase transition from hadron phase to QGP at high temperatures
and at zero baryon chemical potential
is predicted from lattice QCD~\cite{LatticeReview,Lattice05},
and various experimental signals at RHIC suggest the formation of QGP.
On the other hand, 
compressed baryonic matter is created
in heavy-ion collision experiments at lower energies,
and cold and dense baryonic matter is realized
in the core of neutron star.
For these large baryon density matter,
various interesting matter forms have been proposed so far.
These states include
admixture and superfluidity of hyperons
and strange quarks
in neutron star core~\cite{hyperon},
the ${}^3\mathrm{P}_2$ neutron superfluidity~\cite{PTP:PionCondensation},
pion~\cite{pion,PTP:PionCondensation} and kaon~\cite{kaon} condensations,
color ferromagnetic state~\cite{ColorFerro},
color superconductor (CSC)~\cite{CSC},
in addition to the formation of 
baryon rich QGP~\cite{BaryonRichQGP:Nstar,BaryonRichQGP:NNcollision}.

The lattice QCD Monte-Carlo simulations are possible
for hot baryon-free nuclear matter,
and matter at small baryon density~\cite{Muroya2003,LatticeFiniteMu}
can be studied by, for example,
the Taylor expansion method around $\mu=0$~\cite{LatticeFiniteMu,Taylor},
analytic continuation method~\cite{AC},
canonical ensemble method~\cite{Canonical}
and the improved reweighting method~\cite{FodorKatz}.
However, properties of highly compressed matter
are still under debate~\cite{Lattice05,Taylor,AC,Canonical,FodorKatz}.
This is because the fermion determinant, which is used as
the weight in the Monte-Carlo simulation, becomes complex
at finite chemical potential~\cite{MuOnLattice,FiniteMu:Problems}.
Thus, in order to attack the problem of compressed baryonic matter,
it is necessary to invoke some approximations in QCD
or to apply some effective models~\cite{HatsudaKunihiro,Fukushima2004}.
A possible approach is to study color SU(2)
QCD~\cite{Muroya2003,Fukushima2004,SU2,Nishida2004a},
where the fermion determinant is still a real number even at finite 
baryon chemical potential,
but there are several essential differences
between color SU(2) and SU(3) QCD.
For example, the color anti-symmetric diquark pair becomes color singlet,
whose nature would be very different from those discussed in the context of CSC,
and this diquark pair is nothing but a baryon which is a boson 
in color SU(2) QCD.

One of the most instructive approximations to investigate the
finite temperature $T$ and chemical potential $\mu$ of QCD is to consider
the strong coupling limit of lattice
QCD~\cite{Kawamoto1981,KS1981,HKS1982,Kluberg-Stern,DKS1984,DKS1986,DHK1985,Fukushima2004,SU2,Nishida2004a,Azcoiti,Ilgenfritz1985,Faldt1986,Bilic,BilicCleymans,Nishida2004b,XQLuo,MDP,KawamotoShigemoto}.
In fact, effective free energy at finite $T$ 
of strong coupling lattice QCD
was analytically derived and predicted the second order
chiral symmetry restoration temperature~\cite{DKS1984,DKS1986}.
The strong coupling limit lattice QCD effective action for finite
$\mu$ and zero temperature ($T=0$) was also derived
with the help of lattice chemical potential~\cite{MuOnLattice}
and predicted a phase transition near the density of baryonic
formation~\cite{DHK1985}.
These investigations triggered many later analytic~\cite{Fukushima2004,SU2,Nishida2004a,Azcoiti,Ilgenfritz1985,Faldt1986,Bilic,BilicCleymans,Nishida2004b,XQLuo}
and semi-analytic~\cite{MDP}
investigations of finite $T$ and $\mu$ of lattice QCD
in the strong coupling.
It is worth to mention that the Monte-Carlo numerical results of lattice QCD
should reproduce the analytic result of the strong coupling,
and the qualitative nature,
and even quantitative nature for some physical values such as meson masses,
are quite close to the reality of finite coupling
results~\cite{KS1981,KawamotoShigemoto}.

Based on these past experiences,
there have been recently renewed interests of strong coupling
lattice QCD as an instructive guide to finite $T$
and $\mu$ QCD.
The effective free energy of strong coupling limit of lattice QCD action
for SU(2) was analytically derived in \cite{Nishida2004a}.
The effective free energy of finite $\mu$ and zero temperature ($T=0$)
for SU(3) was derived by Azcoiti et al.~\cite{Azcoiti},
who developed a method to decompose the coupling term of the baryon
and three quarks into the coupling terms of diquark auxiliary field 
($\phi_a$) with two quarks and those of $\phi_a$ with a quark and a baryon.
The effective free energy at finite $T$ and $\mu$ for SU(3)
was obtained in \cite{Nishida2004b}, 
but the baryon effects are ignored there.
Thus there is no work which takes account of 
both finite temperature and baryon effects
in the strong coupling limit of lattice QCD for color SU(3) yet.

In this paper, we study the phase diagram of quark matter
at finite temperature ($T$) and finite chemical potential ($\mu$)
in the strong coupling limit of lattice QCD for color SU(3).
We derive an analytical expression of the effective free energy
as a function of $T$ and $\mu$.
We take account of
both the mesonic and baryonic composite terms
in the $1/d$ expansion of the lattice QCD action,
and perform the temporal link variable ($U_0$)
integral exactly in the Polyakov gauge
with anti-periodic boundary condition for fermions,
while we ignore the effects of finite diquark condensate.
Firstly, our treatment is different from the works
by Nishida~\cite{Nishida2004b} and Bilic et al.~\cite{Bilic,BilicCleymans},
who extensively studied the phase diagram
with the leading term in the $1/d$ expansion
containing only the mesonic composites.
Secondly, our formulation is different from the work
by Azcoiti et al.~\cite{Azcoiti},
who made the one link integral also for $U_0$
which is an approximate treatment at finite temperatures.
Thirdly,
we propose a way to include the diquark condensate
as a color singlet order parameter in Subsec.~\ref{Subsec:diquark},
although extensive study is not carried out
and will be reported elsewhere.

This paper is organized as follows.
In Sec.~\ref{Sec:Model},
we derive an analytical expression of the effective free energy
in the strong coupling limit of color SU(3) lattice QCD
with finite temperature and quark chemical potential.
In Sec.~\ref{Sec:Results},
we study the phase diagram of strong coupling limit lattice QCD
in the chiral limit.
In Sec.~\ref{Sec:Discussions},
we examine the parameter dependence of the present model
and compare our results with those in other treatments.
Also we propose a formulation to include diquark condensates
in a mean field ansatz.
We stress the importance of baryon effects in the phase diagram.
We summarize our results in Sec.~\ref{Sec:Summary}.

\section{Effective free energy in the strong coupling limit of lattice QCD}
\label{Sec:Model}

In this section, 
we derive an expression of the effective free energy
in the strong coupling limit lattice QCD with $N_c=3$
with finite temperature and quark chemical potential
in a mean field ansatz
including the baryon effects.
Chemical potential is introduced in the same way
as in Ref.~\cite{MuOnLattice}.
For the finite temperature treatment,
we follow the work by Damgaard, Kawamoto and Shigemoto~\cite{DKS1984,DKS1986},
in which the anti-periodic boundary condition for fermions is exactly treated
and the integral over the temporal link variable $U_0$ is performed
exactly in the Polyakov gauge.
In order to apply this technique,
we have to obtain the effective action in the bilinear form of the quark field.
Such effective actions have been
derived~\cite{DKS1984,DKS1986,Ilgenfritz1985,Faldt1986,Bilic,BilicCleymans,Nishida2004b}
only with the leading order mesonic composite term in the $1/d$ expansion
for color SU(3).
We utilize the idea proposed by Azcoiti et al.~\cite{Azcoiti}
to decompose the baryonic composite term.
Throughout the paper,
both of the temporal and spatial direction points on the lattice,
$\beta=1/T$ and $L$, are assumed to be even integers,
and the lattice spacing is set to be unity.
While $T=1/\beta$ takes discrete values,
the effective free energy is given as a function of $T$ (and $\mu$),
then we consider $T$ as a continuous valued temperature.

\subsection{Strong coupling limit and integral over spatial link variables}
We start from an expression of lattice QCD action
with one species of staggered fermion
for color SU($N_c$).
In the strong coupling limit ($g \to \infty$),
we can ignore the pure gluonic part of the action,
since it is proportional to $1/g^2$.
As a result, the lattice action contains
only those terms including fermions, $S_F$.
\beqar
S_F[U,\chi,\chibar]&=&
S_F^{(U_0)}[U_0,\chi,\chibar]\nonumber\\
&+&\sum_{j=1}^d S_F^{(U_j)}[U_j,\chi,\chibar]
+S_F^{(m)}[\chi,\chibar]\ ,
\\
S_F^{(U_0)}
&=&
 \frac12 \sum_{x}
	\left[
		 \chibar(x)e^{\mu}U_0(x)\chi(x+\hat{0})
	\right.
\nonumber\\
&&~~~
	\left.
		-\chibar(x+\hat{0})e^{-\mu}U^\dagger_0(x)\chi(x)
	\right]\ ,
\\
S_F^{(U_j)}
&=&
 \frac12 \sum_{x} \eta_j(x)
	\left[
		 \chibar(x)U_j(x)\chi(x+\hat{j})
	\right.
\nonumber\\
&&~~~
	\left.
		-\chibar(x+\hat{j})U^\dagger_j(x)\chi(x)
	\right]
\ ,
\\
S_F^{(m)}
&=&m_0\sum_{x}\chibar^a(x)\chi^a(x)
\ ,
\eeqar
where
we introduce the chemical potential $\mu$
in the same way to Ref.~\cite{MuOnLattice}. And
$\eta_j(x)=(-1)^{x_0+x_1+\cdots+x_{j-1}},~(j=1,2,3)$
is a Kogut-Susskind factor.
The staggered fermion $\chi$ represents the quark field,
and the SU($N_c$) matrix $U_\mu$ represents the gauge link variable.

In the first step,
we perform the group integral for spatial link variables,
$U_j(x)$ ($j=1,2,3$).
Integral of the leading and next-to-leading order terms
in the $1/d$ expansion
leads to the following action,
\beqar
&&
\int \Fint{U_j} e^{-\sum_{j=1}^d S_F^{(U_j)}[U,\chi,\chibar]}
 \simeq e^{-S_F^{(j)}}
\ ,\\
\label{Eq:SFj}
&&S_F^{(j)}[\chi^a,\chibar^a]
=
		-\frac12 (M,V_M M)
		-(\bar{B},V_B B)
\ .
\eeqar
where the inner product of fields are defined as
$(A, V B) \equiv \sum_{x,y} A(x) V(x,y) B(y)$.
The mesonic and baryonic composites and their propagators are
defined~\cite{KS1981,HKS1982}
as
\beqar
M(x) &=& \chibar^a(x)\chi^a(x)
\ ,\\
B(x) &=& {1\over N_c!} \eps_{ab\cdots c} \chi^a(x)\chi^b(x) \cdots \chi^c(x)
\ ,\\
\bar{B}(x)
	&=& {1\over N_c!} \eps_{ab\cdots c}
		\chibar^c(x) \cdots \chibar^b(x) \chibar^a(x)
\ ,\\
V_M(x,y)&=&{1\over 4N_c}
\sum_{j=1}^d
\left(\delta_{y,x+\hat{j}} + \delta_{y,x-\hat{j}}\right)
\ ,\\
V_B(x,y)&=&
(-1)^{N_c(N_c-1)/2}
\sum_{j=1}^d \left\{\eta_j(x)\over 2\right\}^{N_c}
\nonumber\\
&&\times
\left(\delta_{y,x+\hat{j}} +(-1)^{N_c} \delta_{y,x-\hat{j}}\right)
\ .
\eeqar
Here we have utilized the SU($N_c$) group integral formulae,
\beqar
&& \int d[U] U_{ab} U^\dagger_{cd} = {1\over N_c} \delta_{ad}\delta_{bc}
\ ,\\
&& \int d[U] U_{ab}U_{cd} \cdots U_{ef}
	= {1\over N_c!} \eps_{ac\cdots e} \eps_{bd\cdots f}
\ .
\eeqar

The baryonic composite action $(\bar{B},V_B B)$
is often ignored with $N_c \geq 3$,
since it is proportional to $1/\sqrt{d^{N_c-2}}$
in the $1/d$ expansion~\cite{Kluberg-Stern}.
This scaling can be understood as follows.
Mesonic and baryonic propagators contains the sum over $j=1, 2, \ldots d$,
and they are considered to be proportional to $d$.
In order to keep the mesonic term $(M,V_M M)/2$ finite in the large $d$ limit, 
the mesonic composite should be proportional to $d^{-1/2}$.
Then the quark field, the baryonic composite, and the baryonic composite action
are proportional to $d^{-1/4}$, $d^{-N_c/4}$, and $d^{-(N_c-2)/2}$,
respectively.
For the discussion of dense baryonic matter, however,
we expect larger baryon effects.
Thus we keep this baryonic composite action and proceed.
In the following discussion, we consider $N_c=3$ case.

\subsection{Auxiliary fields}

The effective action Eq.~(\ref{Eq:SFj}) contains six fermion terms,
while we have to obtain the effective action in the bilinear
or pfaffian~\cite{ZinnJustin}
form in $\chi$ to perform the quark field integral at finite temperature.
In order to reduce the power in $\chi$ and $\chibar$, 
we introduce several auxiliary fields.

The highest power term $\bar{B}B$ containing six quarks
can be reduced by introducing the auxiliary baryon field $b$
through the following identity,
\beq
\label{Eq:BB}
e^{(\bar{B}, V_B B)}
=\mbox{det}V_B \int \Fint{\bar{b},b}
	e^{
		-(\bar{b}, V_B^{-1} b)
		+(\bar{b}, B)+(\bar{B}, b)
	}
\ .
\eeq

Next, we decompose the coupling terms of the baryon and three quarks
by using the technique developed in \cite{Azcoiti}.
We consider the following composite diquark field $D_a$.
\beq
D_a = {\gamma\over2} \eps_{abc}\chi^b \chi^c
	+ {1\over3\gamma}\chibar^a b
\ ,\quad
D_a^\dagger = {\gamma\over2} \eps_{abc}\chibar^c \chibar^b
	+ {1\over3\gamma}\bar{b}\chi^a
\ .\label{Eq:Diquark}
\eeq
These are the combinations of
diquark and baryon-antiquark (antibaryon-quark) pairs,
and have the color transformation properties of
$\bar{\bold{3}}$ and $\bold{3}$ for $D_a$ and $D_a^\dagger$, respectively.
The parameter $\gamma$ is introduced so as to generate
the coupling terms, $\bar{B}b+\bar{b}B$.
\beqar
D_a^\dagger D_a &=&\bar{B} b +\bar{b} B + Y\ ,\\
Y &=& {\gamma^2\over2} M^2 -{1\over9\gamma^2} M \bar{b}b\ .
\label{Eq:DDY}
\eeqar
The decomposition in Ref.~\cite{Azcoiti} corresponds to $\gamma=2$.
The product $D_a^\dagger D_a$ can be generated
by the auxiliary field $\phi_a$, and we can replace $\barb B +\bar{B}b$
terms as follows.
\beq
e^{
	\bar{b}B + \bar{B}b
	}
=
\int d[\phi_a, \phi^\dagger_a]
	e^{
		-\phi_a^\dagger \phi_a
		+
		(\phi_a^\dagger D_a+D_a^\dagger \phi_a)
		- Y
	}
\ ,
\eeq
where the expectation value of $\phi_a$
is the same as that for $D_a$, 
$\langle\phi_a\rangle = \langle{D_a}\rangle$.

In terms of the $1/d$ expansion,
the baryonic auxiliary field $b$ is proportional to $d^{1/4}$,
provided that the exponent in Eq.~(\ref{Eq:BB}) is ${\cal O}(d^{-1/2})$.
Thus the second term $\chibar b$ in $D_a$ is ${\cal O}(1)$,
while the first term $\eps_{abc}\chi^b\chi^c$ is proportional to $d^{-1/2}$,
and we expect the dominance of the second term for large $d$.
This may be the reason why we need the baryon-antiquark pair
in discussing the diquark pair condensate.

In the next step, we decompose the coupling term 
of the baryon and mesonic composite, $M\bar{b}b$, 
by introducing the baryon potential auxiliary field $\omega$
though the identity,
\beq
e^{M \bar{b}b/9\gamma^2}=\int d[\omega]\,e^{
		-\omega^2/2-\omega(\alpha M+g_\omega\bar{b}b)
		-\alpha^2 M^2/2
		}
\ ,
\eeq
where $g_\omega=1/9\alpha\gamma^2$ and 
$\VEV{\omega}=-\VEV{\alpha M+g_\omega\barb{b}}$.
Note that the local four baryon term becomes zero,
$\bar{b}(x)b(x)\bar{b}(x)b(x)=0$,
due to the Grassmann variable nature,
which is a natural consequence of staggered fermion formulation for one-flavor.

Finally, we introduce the auxiliary field for chiral condensate.
It is interesting to note that
we have additional ``mass'' terms, $(\gamma^2+\alpha^2)M^2/2$
for the mesonic composite $M$
through the decomposition of baryonic composite action
by introducing the auxiliary diquark and baryon potential fields.
These terms are made of four quarks,
and it is not easy to handle in the quark integral.
Therefore, we include these terms in the hopping term,
\beq
\frac12(M,V_M M)-\frac12(\gamma^2+\alpha^2)M^2
= \frac12 (M,\VM M)
\ ,
\eeq
then it becomes possible to bosonize as,
\beqar
&&
e^{\frac12 (M,\VM M)}=\int\Fint{\sigM}
	e^{
		-\frac12 (\sigM, \VM \sigM)
		- (\sigM, \VM M)
		}
\ ,\label{Eq:IntegM}\\
&&
\VM(x,y)=V_M(x,y)
	-(\gamma^2+\alpha^2)\delta_{x,y}
\ .
\eeqar
The expectation value of $\sigM$ is given as
$\VEV{\sigM}=-\VEV{M}$.
The mesonic propagator $\VM$ has negative eigenvalues
as well as positive ones,
and thus it is expected that instability is introduced
in the gaussian integration.
However in the mean field ansatz,
vacuum expectation value of the meson is 
introduced so that next neighboring $(x,x+\hat{\mu})$
dependence is suppressed,
which corresponds to the suppression of $(x,y)$ dependence in $\VM(x,y)$.
In this way, we circumvent the instability,
which is done in the literature and we show in the following.

After these sequential introduction of auxiliary fields, 
we obtain the action of quarks, baryons, diquarks, 
baryon potential, and the chiral condensate as follows.
\beqar
&&
S_F=S_F^{(X)}+S_F^{(q)}
\ ,\\
&&
S_F^{(X)}[b,\bar{b},\phi,\phi^\dagger,\sigM,\omega]
\nonumber\\
&&~=
 (\bar{b},\VB^{-1} b)
+(\phi^\dagger,\phi)
+\frac12 (\omega,\omega)
+\frac12 (\sigM, \VM\sigM)
\ ,
\\
&&
S_F^{(q)}[U_0,\chi,\chibar,b,\bar{b},\phi,\phi^\dagger,\sigM,\omega]
\nonumber\\
&&~=S_F^{U_0}+(\mq,M)
+{1\over3\gamma}\left[
	 (\chibar^a, \phi_a^\dagger b)
	+(\bar{b}\phi_a,\chi^a)
	\right]
\nonumber\\
&&~~~~
+\frac{\gamma}{2}\eps_{cab}\left[
	 (\phi_c^\dagger, \chi^a \chi^b)
	+(\chibar^b\chibar^a,\phi_c)
	\right]
\ ,
\label{Eq:SFQ-A}
\\
&&
\mq = \VM \sigM+ \alpha\omega + m_0
\label{Eq:SIG_Q}
\ .
\eeqar
where $S_F^{(X)}$ and $S_F^{(q)}$ are
the action of pure auxiliary fields
and the action containing quarks, respectively.
The inverse baryonic propagator is modified as
\beq
\VB^{-1}(x,y)= V_B^{-1}(x,y)+g_\omega\omega \delta_{x,y}
\ ,\quad
g_\omega = {1\over9\alpha\gamma^2}
\ .
\eeq

It is noteworthy that the quark action is
decomposed into that for each spatial point,
$\bold{x}$.
Therefore, it would be good enough to assume that
bosonic auxiliary fields have constant values,
i.e. mean field ansatz would be valid.
This simplifies the term
containing $\VM \sigM$ as follows,
\beqar
&&
\VM\sigM
= \asig \sigM
\ ,\quad
\frac12(\sigM,\VM\sigM)
= \frac{\beta L^3}{2} \asig\sigM^2
\ ,\\
&&
\asig
={d\over2N_c}-(\gamma^2+\alpha^2)
\ .
\eeqar

\subsection{Quark integral}

In order to perform the quark integral,
we would like to separate the action into terms,
each of which has as small number of quark fields as possible.
In the quark action Eq.~(\ref{Eq:SFQ-A}),
the time component of the link variable $U_0$ connect
the quark field of different (imaginary) time,
and all the quark fields with different times couple.

This coupling is known to be separated by using
the Fourier transformation for the fermion fields~\cite{DKS1984,DKS1986},
\beqar
&&
\psi(x) = {1\over \sqrt{\beta}}\sum_{m=1}^\beta
		e^{ik_m \tau} \psi_m(\bold{x})
\ ,\\
&&
\bar{\psi}(x) = {1\over \sqrt{\beta}}\sum_{m=1}^\beta
		e^{-ik_m \tau} \bar{\psi}_m(\bold{x})
\ ,
\eeqar
where $\psi$ stands for $\chi$ or $b$,
and the Matsubara frequencies, $k_m=2\pi(m - 1/2)/\beta$,
are selected to satisfy anti-periodic condition of fermions,
$\psi(\beta,\bold{x})=-\psi(0,\bold{x})$,
to introduce temperature effects.

We ignore the time dependence of bosonic auxiliary fields, 
$\phi, \phi^\dagger, \omega, \sigM$
(static approximation),
and we work in the Polyakov gauge,
where the link variable $U_0$ is diagonal and independent on time,
\beq
U_0(\bold{x},\tau)
= {\rm diag}(	e^{i\theta_1(\bold{x})/\beta},
		e^{i\theta_2(\bold{x})/\beta},
		e^{i\theta_3(\bold{x})/\beta})
\ ,
\eeq
with the condition $\theta_1+\theta_2+\theta_3 = 0$.
The quark action is found to be represented
in the form of pfaffian,
\beqar
S_F^{(q)}
&=&
\frac12 
\sum_{\bold{x},m,n}
 		(\chibar^a_m, \chi^a_m)
		\bold{G}^{-1}_{ab}(m,n)
		\begin{pmatrix}
		\chi^b_n\\
		\chibar^b_n\\
		\end{pmatrix}
\nonumber\\
&+&\sum_{\bold{x},m}
		\left(
		 \bar{C}^a_m \chi^a_m
		+\chibar^a_m C^a_m
		\right)
\ ,
\label{Eq:SFQ-B}
\eeqar
\beq
\bold{G}^{-1}_{ab}(m,n;\theta_a)=
		\begin{pmatrix}
		  B^a(k_m) \delta_{ab}\delta_{mn}
		& -\gamma\eps_{cab}\phi_c \delta'_{mn}\\
		   \gamma\eps_{cab}\phi_c^\dagger \delta'_{mn}
		& - B^a(k_m) \delta_{ab}\delta_{mn}\\
		\end{pmatrix}
\ ,
\eeq
\beq
B^a(k) = \mq 
	+ i \sin(k+\theta^a/\beta-i\mu)
\ ,
\eeq
\beq
C^a_m = \frac{1}{3\gamma}\phi_a^\dagger b_m
\ ,\quad
\bar{C}^a_m = \frac{1}{3\gamma}\bar{b}_m\phi_a
\ .
\eeq
In the first line of Eq.~(\ref{Eq:SFQ-B}), we have used the notation
$\delta'_{mn}=\delta_{m,\beta-n+1}$.
The second term in Eq.~(\ref{Eq:SFQ-B}) can be absorbed into the first term
by shifting the quark field at a cost of
producing another term $S_F^{(C)}$, which is bilinear in $b$ and $\barb$.
\beqar
S_F^{(q)}
&=&
 \frac12 
 \sum_{\bold{x},m,n}
 		(\chibar^a_m, \chi^a_m)
		\bold{G}^{-1}_{ab}(m,n)
		\begin{pmatrix}
		\chi^b_n\\
		\chibar^b_n\\
		\end{pmatrix}
 +S_F^{(C)}
\ ,
\\
\label{Eq:SFC}
S_F^{(C)}
&=&
-\frac12 
 \sum_{\bold{x},m,n}
 		(\bar{C}^a_m, -C^a_m)
		\bold{G}_{ab}(m,n)
		\begin{pmatrix}
		C^b_n\\
		-\bar{C}^b_n\\
		\end{pmatrix}
\ .
\eeqar
The action $S_F^{(C)}$ appears from 
the baryon-quark coupling generated by the diquark condensate,
and it is very difficult to handle with finite values of $\phi$.
Another treatment to replace this coupling by other terms
will be discussed in Subsec.~\ref{Subsec:diquark},
and we temporarily ignore $S_F^{(C)}$ here. In this case,
by symmetrizing for $m$ and $m'=\beta - m + 1$ in Eq.~(\ref{Eq:SFQ-B}),
We obtain the block diagonal form of $S_F^q$,
\begin{equation}
\label{Eq:SFq}
S_F^{(q)}=
 \sum_{\bold{x},a,b}\sum_{m=1}^{\beta/2}
 		(\chibar^a_m,\chi^a_{m'})
		\bold{g}_{ab}(k_m)
		\begin{pmatrix}
		\chi^b_m	\\
		-\chibar^b_{m'}\\
		\end{pmatrix}\ .
\end{equation}
\begin{equation}
\label{Eq:gdet}
\bold{g}_{ab}(k)
=
		\begin{pmatrix}
		  \bold{B}(k)_{ab} & \bold{A}_{ab} \\
		  \bold{A}_{ab}^*  & \bold{B}(-k)_{ab} \\
		\end{pmatrix}
\ ,
\end{equation}
Here, $(\bold{B}(k))_{ab}=\delta_{ab} B^a(k)$,
$(\bold{A})_{ab} = \gamma\eps_{cab}\phi_c$ and
we have used the relation,
$k_{m'} = 2\pi - k_m = -k_m ({\rm mod}(2\pi))$.
Since $\chibar_m,~\chi_{m'}$ are independent of each other,
Grassmann integral over $\chi, \chibar$ leads to a determinant:
\beq
\int \Fint{\chi,\chibar}e^{-S_F^{(q)}}
=
	\prod_{\bold{x}}G(\bold{x})
\ .
\eeq
\beq
\label{Eq:GdetA}
G(\bold{x})
	\equiv\prod_{m=1}^{\beta/2}\mbox{det}\left[\bold{g}_{ab}(k_m)\right]
	=\prod_{m=1}^{\beta}\mbox{det}\left[\bold{g}_{ab}(k_m)\right]^{1/2}
\ .
\eeq
The $G(\bold{x})$ is evaluated
by the direct calculation of $\det$:
\begin{eqnarray}
G(\bold{x})&=&\prod_{m=1}^{\beta/2}\Bigl[
\gamma^4(B_1|\phi_1|^2 +B_2|\phi_2|^2 +B_3|\phi_3|^2 )
\nonumber\\
&&\times(B_1'|\phi_1|^2+B_2'|\phi_2|^2+B_3'|\phi_3|^2)
\nonumber\\
&+&\gamma^2\sum_{(a,b,c)=cyc.} B_a B_a' |\phi_a|^2 (B_b B_c' + B_b' B_c)
\nonumber\\
&+&B_1 B_2 B_3 B_1' B_2' B_3'\Bigr]\ , \label{Eq:detg}
\end{eqnarray}
where $B_a = B^a(k_m), B_a'=B^a(-k_m)$.
In a similar way to that in Ref. \cite{Nishida2004a},
we can perform the Matsubara frequency product $\prod_m$,
\begin{equation}
\label{Eq:GdetB}
G(\bold{x})=\prod_j\bigl[1 + e^{-i\beta z_j(\bold{x})} \bigr]^{1/2},
\end{equation}
where
$z_j(\bold{x})$ is the solution of
$\mbox{det}\left[\bold{g}_{ab}(k_m)\right]=0$.
The explicit derivation is given in the Appendix~\ref{App:Sum}.

\subsection{Effective free energy at zero diquark condensate}
When the diquark condensate is zero, $\phi_a=0$,
we know the solutions of $\mbox{det}\left[\bold{g}_{ab}(k_m)\right]=0$.
We can take $B_a=B^a(k)=0$ and $B_a'=B^a(-k)=0$,
and we get four solutions for each $a$,
\beq
	i\beta z_a = \pm\left[
		i(\theta_a -i \beta\mu) \pm \beta\Eq(\mq)
	\right]
\ ,
\eeq
where $E_q(\mq)=\arcsinh \mq$
is one dimensional quark excitation energy.
We can then explicitly write the quark integral results as,
\beqar
G(\bold{x})
&=& \prod_{a=1}^{N_c}
	\left[
		(1+e^{-i(\theta_a-i\beta\mu)-\beta\Eq})
		(1+e^{-i(\theta_a-i\beta\mu)+\beta\Eq})
	\right.
\nonumber\\
&&
	\times
	\left.
		(1+e^{ i(\theta_a-i\beta\mu)-\beta\Eq})
		(1+e^{ i(\theta_a-i\beta\mu)+\beta\Eq})
	\right]^{1/2}
\nonumber\\
&=& \prod_{a=1}^{N_c}
		2\left(\cos(\theta-i\beta\mu)+\cosh\Eq\right)
\nonumber\\
&=& \prod_a 2\left(C_\sigM + C_\mu \cos\theta_a + i S_\mu \sin\theta_a\right)
\label{Eq:Gx}
\ ,
\\
C_\sigM &=&\cosh\beta\Eq
\ ,\;
C_\mu =  \cosh\beta\mu
\ ,\;
S_\mu =  \sinh\,\beta\mu
\ .
\eeqar

There are three comments on the phase of $G(\bold{x})$ in order.
First, the square root in the quark determinant (Eq.~(\ref{Eq:GdetB}))
is the Pfaffian root~\cite{ZinnJustin},
but it comes from the square root in Eq.~(\ref{Eq:GdetA}),
where we extend the range of the Matsubara product
from $\beta/2$ to $\beta$ by using the even function nature of
$\mbox{det}\left[\bold{g}_{ab}(k_m)\right]
=\mbox{det}\left[\bold{g}_{ab}(-k_m=k_{m'})\right]$.
Since the phase of the square root in Eq.~(\ref{Eq:GdetA})
is defined to reproduce the product up to $\beta/2$,
there is no phase ambiguity.
This is because 
we can represent the quark action $S_F^{(q)}$ in Eq.~(\ref{Eq:SFq})
in a usual bilinear fermion action by defining a new fermion field as
$(\psi^b_{m'},\bar{\psi}^a_{m'})\equiv(-\chibar^b_{m'},\chi^a_{m'})$,
then it is not necessary to introduce the Pfaffian root.
Secondly, we may have a phase coming from a constant in $\log G(\bold{x})$
as shown in the Appendix \ref{App:Sum},
but this constant does not depend on the gluon configurations
$\theta_a(\bold{x})$.
As a result, 
we have a fixed phase for a given spatial point $\bold{x}$, and
we get well-defined integral of $G(\bold{x})$ over $dU_0(\bold{x})$.
In this way, we expect that we get reasonable continuum limit,
which we can still consider in the strong coupling region.
Thirdly, in order to take one flavor fermion configuration,
we have to take one quarter root of Eq.~(\ref{Eq:Gx}),
where the well known phase ambiguity of
complex number appears~\cite{Golterman2006}.
Here, we simply consider configuration with one species of staggered fermion,
which in general regarded as four flavor configurations,
and do not take into account the complexity of the four flavor feature
of staggered fermion.

By using the SU(3) Haar measure in the Polyakov gauge,
\beqar
&&
\int dU_0(\bold{x}) =
	\int{d\theta_1 d\theta_2 d\theta_3 \over (2\pi)^3}
	\Delta(\theta_1,\theta_2,\theta_3)
\ ,\\
&&
\Delta =\delta(\theta_1+\theta_2+\theta_3)
	\prod_{i<j}
	(1-\cos(\theta_i-\theta_j))
\ ,
\eeqar
the integral over $U_0$ can be performed analytically.
\beqar
&&
\exp\left(-\beta L^3 \FeffQ\right)
\equiv 
\prod_{\bold{x}}
\int dU_0(\bold{x}) G(\bold{x})
\ ,\\
&&
\FeffQ(\sigM)
=
-T\log\left[
	\frac43\left(
	C_\sigM^3 -\frac12 C_\sigM +\frac14 C_{N_c\mu}
	\right)
	\right]
\ ,\\
&&
C_\sigM = \cosh{\Eq\over T}
\ ,\quad
C_{N_c\mu} = \cosh{N_c\mu\over T}
\ .
\eeqar
where $T=1/\beta$ is regarded as a temperature.
It is interesting to find that $G(\bold{x})$ can have a complex phase
for a given gluon configuration, but after integration over the temporal
link variables, we have a positive definite results
and we do not have the sign problem.
This is one of the merits in the strong coupling limit
in which the link integral can be performed in an analytic manner.
This result is consistent with that for SU($N_c$)
shown in, for example, Ref.~\cite{Nishida2004b},
\beq
\FeffQ=-T\log\left\{{\sinh[(N_c+1)E/T]\over\sinh[E/T]}+2C_{N_c\mu}\right\}
\ ,
\eeq
while $C_{N_c\mu}$ term does not appear
in $\textrm{U}(N_c)$ case~\cite{DKS1984,DKS1986}.

When the diquark condensate is zero, $\phi_a=0$,
we can ignore $S_F^{(C)}$, and it becomes possible to perform
baryon integral, too:
\begin{equation*}
\FeffB(g_{\omega}\omega)
={1\over\beta L^3}\log{\rm Det}\left[
		1 + g_{\omega}\omega V_B
		\right]\ .
\end{equation*}
As shown in the Appendix~\ref{App:Baryon},
we can evaluate 
this determinant by using the Fourier transformation.
For large spatial lattice size $L$,
by replacing the sum over $\bold{k}$ by the integral,
we get the following expression,
\begin{eqnarray}
\FeffB(g_{\omega}\omega)
&\simeq&
	\frac{-a^{(b)}_0/2}{\left(4\pi\Lambda^3/3\right)}
	\int_0^{\Lambda} 4\pi k^2 dk 
	\log\left[1 + {g_{\omega}^2\omega^2 k^2\over16}\right]
\nonumber\\
&&=
	-a_0^{(b)} f^{(b)}\left(\frac{g_{\omega}\omega\Lambda}{4}\right)
\label{bdet}\ ,
\end{eqnarray}
where, $a^{(b)}_0 = 1.0055\ ,~\Lambda =  1.01502 \times \pi/2$,
and $f^{(b)}(x)$ is given as,
\begin{equation}
f^{(b)}(x)
=\frac12 \log(1+x^2)
	- {1\over x^3}\left[
		\arctan{x} - x + {x^3\over 3}
	\right]
	\ .
\end{equation}
Since this baryon determinant term $\FeffB$ is independent
from $T$ and $\mu$, it would be more convenient for the later discussion
to separate the quadratic term in $\omega$ as follows,
\beqar
{\omega^2\over2}+\FeffB(g_{\omega}\omega)
&=&\frac12 a_{\omega} \omega^2 + \Delta\FeffB(g_{\omega}\omega)
\ ,
\\
a_{\omega}
&=& 1 -{3\over5}a_0^{(b)}\left({g_{\omega}\Lambda\over4}\right)^2
\ ,
\eeqar
where $\Delta \FeffB = {\cal O}(\omega^4)$
at small $\omega$ values.
The second term in $a_{\omega}$ comes from the expansion of Eq.(\ref{bdet}).

After quark, gluon and baryon integral, 
the total effective free energy is obtained as,
\beq
\label{Eq:FeffOrg}
\FEFF
=
	 \frac12 \asig \sigM^2
	+\frac12 a_{\omega} \omega^2
	+ \FeffQ(\mq)
	+ \Delta\FeffB(g_{\omega}\omega)
\ ,
\eeq
where $\mq=\asig\sigM+\alpha\omega+m_0$.

\subsection{Stability and equilibrium condition}

\begin{figure}
\Psfig{9cm}{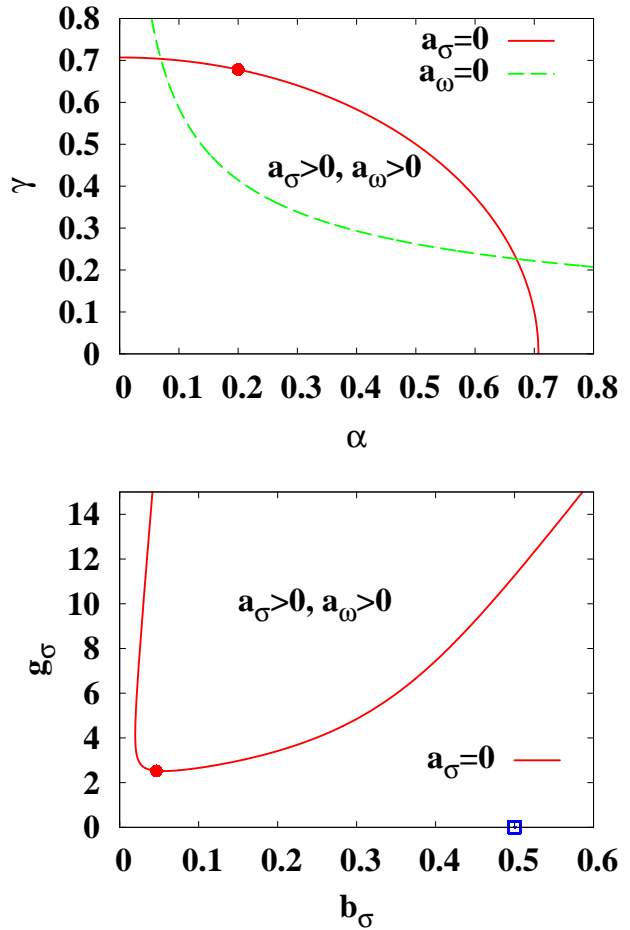}{}
\caption{
(Color online)
Parameter range which satisfies 
the conditions of stability and high $T$ chiral symmetry restoration.
The solid dot represents the parameter set,
$\alpha=0.2, \alpha^2+\gamma^2 =1/2-0$,
which we adopt in the later discussion, 
and the square shows the parameter set
without baryon effects.
}\label{Fig:ParRange}
\end{figure}

We have introduced two parameters, $\gamma$ and $\alpha$,
and 
two auxiliary fields, $\sigM$ and $\omega$,
in the derivation of the effective free energy, Eq.~(\ref{Eq:FeffOrg}).
Since we have introduced these parameters and fields through identities,
the final results should not depend on these parameters
if all the integrals are completed.
However, we are working in the mean field ansatz,
then we may have some parameter dependence.
In principle, we should select the parameters 
so as to keep the mean field ansatz valid;
the effective free energy should be stable against the variation of
the fields, $\sigM$ and $\omega$,
and the effective free energy (the free-energy density)
at equilibrium should be stationary against the variation of parameters.
We further require that 
the chiral symmetry is restored at very high temperatures.
The stability of the effective free energy
against the variation of $\sigM$ is satisfied when $\asig > 0$,
and the chiral symmetry restoration at very high temperatures
is ensured when $a_\omega > 0$.
The region of $\alpha$ and $\gamma$ which satisfies both of the conditions
are shown in the upper panel of Fig.~\ref{Fig:ParRange}.
The parameter dependence on parameters are
discussed in Sec.~\ref{Subsec:pardep}.

By using the equilibrium condition, 
two auxiliary fields are related to each other,
then we can obtain the effective free energy as a function of 
one order parameter.
At equilibrium, the effective free energy is stationary
with respect to $\sigM$ and $\omega$,
\beqar
{\partial\FEFF\over\partial\sigM}
&=& \asig\sigM+\asig{\partial\FeffQ\over\partial\mq}=0
\ ,\\
{\partial\FEFF\over\partial\omega}
&=& a_\omega\omega
+\alpha{\partial\FeffQ\over\partial\mq}
+{\partial\Delta\FeffB\over\partial\omega}=0
\ .
\eeqar
The effects of $\Delta\FeffB$ is small when the fields are small,
then in this case $\omega$ can be represented
by the chiral condensate $\sigM$ as,
\beqar
\label{Eq:Linear}
{a_\omega\over\alpha}\omega
\simeq \sigM
=-{\partial\FeffQ\over\partial\mq}
=-{\partial\FEFF\over\partial m_0}
\ .
\eeqar
With this approximation for $\omega$,
the effective free energy is given as a function of $\sigM$ as,
\beqar
\label{Eq:Feff}
\FEFF &=& \frac{\bsig}{2}\sigM^2 + \FeffQ(\mq)
	+\Delta\FeffB(\gsig \sigM)
\ ,\\
\mq&=&\bsig\sigM+m_0
\ ,\\
\label{Eq:bsiggsig}
\bsig&=& \asig+{\alpha^2\over a_\omega}
\ ,\quad
\gsig = {\alpha g_\omega \over a_\omega}
\ .
\eeqar
With this form of the effective free energy,
the meaning of parameters are a little more clear.
The constituent quark mass is a linear function of $\sigM$,
then the coefficient $\bsig$ represents the polarizability
of the chiral condensate, which is modified by the baryonic composite effects.
The parameter $\gsig$ determines the strength
of the coupling of the chiral condensate and the baryon,
and $\Delta\FeffB(\gsig\sigM)$ represents
the repulsive self-interaction of $\sigM$ 
coming from the baryon integral.

The parameters $\bsig$ and $\gsig$
are related to $\alpha$ and $\gamma$,
and they have the region which satisfies the conditions
of stability and high $T$ chiral symmetry restoration
as shown in the lower panel of Fig.~\ref{Fig:ParRange}.
We notice $\Delta\FeffB(\gsig\sigM)$ has the positive
value for any $\sigM$, hence, the smaller $\gsig$ leads to smaller
$\mathcal{F}_\mathrm{eff}$. 
Thus, we choose those parameters to give a small coupling $\gsig$
for a given polarizability $\bsig$.
The smallest $\gsig$ at a fixed $\bsig$ is obtained
in the limit of $\asig \to +0$,
or $\gamma^2 + \alpha^2 \to 1/2 - 0$.
There is no singular behavior in the effective free energy in this limit
as a function of the chiral condensate $\sigM$ in Eq.~(\ref{Eq:Feff}).
For numerical calculations,
we adopt $\alpha=0.2$,
which gives almost the smallest $\gsig$
as shown by the filled circle in Fig.~\ref{Fig:ParRange},
as a typical value in the later discussion.

\section{Phase Structure}
\label{Sec:Results}

In the previous section, we have demonstrated
that the effective action in the strong coupling limit lattice QCD
can be obtained in an almost analytic way
with $1/d$ expansion and mean field ansatz,
when the diquark condensate is zero.
Especially, we focus our attention to the chiral phase transition.

In this section,
we discuss the phase diagram
based on the effective free energy 
as a function of the chiral condensate in Eq.~(\ref{Eq:Feff})
in the chiral limit, $m_0=0$.
Since we utilize the linear approximation
($a_\omega\omega/\alpha\simeq\sigM$, see Eq.~(\ref{Eq:Linear})),
equilibrium value of $\sigM$ slightly differs from that of 
$-\partial\FEFF/\partial m_0=-\langle\bar{\chi}\chi\rangle$. 
However the difference of those is small and only by around 1 \%
as discussed in Subsec.~\ref{Subsec:phase}.
For numerical calculations,
we adopt a parameter choice of $\alpha=0.2$
and $\alpha^2+\gamma^2 \to 1/2$,
which gives $\bsig\simeq0.0465$
and $\gsig\simeq2.527$.
This value of $\gsig$ is almost the smallest value allowed in this model.

\begin{figure}
\Psfig{9cm}{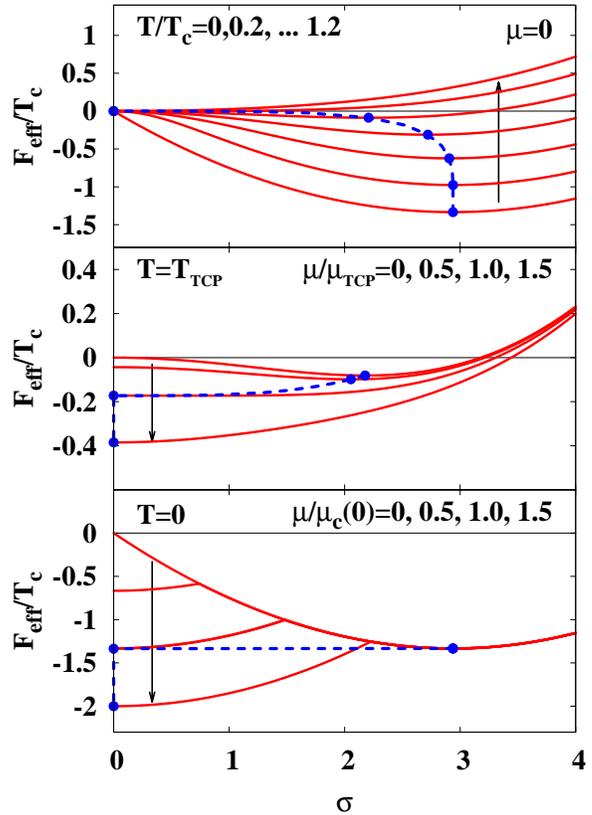}{}
\caption{
(Color online)
Effective potential as a function of $\sigM$.
Upper, middle, and lower panels show
the effective free energies $\FEFF$
in the unit of $T_c$
at $\mu=0$, $T=T_\mathrm{TCP}$, and $T=0$, respectively.
Dots represent the equilibrium points,
and dashed lines connect these points.
}\label{Fig:Esurf}
\end{figure}

\subsection{Zero temperature}

It would be instructive to analyze several limits
of the effective free energy Eq.~(\ref{Eq:Feff}).
The effective free energy from the quark integral
$\FeffQ$ depends on the temperature and chemical potential.
At zero temperature,
$\FeffQ$ can be reduced to
\begin{align}
\FeffQ(\mq;T \to 0,\mu)=
	\begin{cases}
	- N_c\, \Eq & (\Eq > \mu)\ , \\
	- N_c\,\mu & (\Eq < \mu)\ , \\
	\end{cases}
\end{align}
where $\Eq=\arcsinh m_q$ is a quark excitation energy.
In vacuum in the chiral limit, $(T,\mu,m_0)=(0,0,0)$,
$\FeffQ$ has a linear term in $\sigM$,
while other parts of the effective free energy start with $\sigM^2$,
then we necessarily have a finite equilibrium value of $\sigM$.
On the other hand, for a finite chemical potential, 
$\FeffQ$ becomes independent from $\sigM$ for small $\sigM$,
and the effective free energy start
with the quadratic term, $\bsig\sigM^2$ in the chiral limit.
Then we have a local minimum at $\sigM=0$ when $\mu$ is finite
even if it is very small,
as shown in the bottom panel of Fig.~\ref{Fig:Esurf}.
As a result, the first order chiral phase transition
occurs at the chemical potential which satisfies,
\beq
-3\mu_c^\mathrm{(1st)}(T=0) = \FEFF(\sigM_{0};T=0,\mu=0)
\ ,
\eeq
where $\sigM_0$ stands for the vacuum equilibrium value of $\sigM$.

\subsection{Small chemical potentials}

\begin{figure}
\Psfig{9cm}{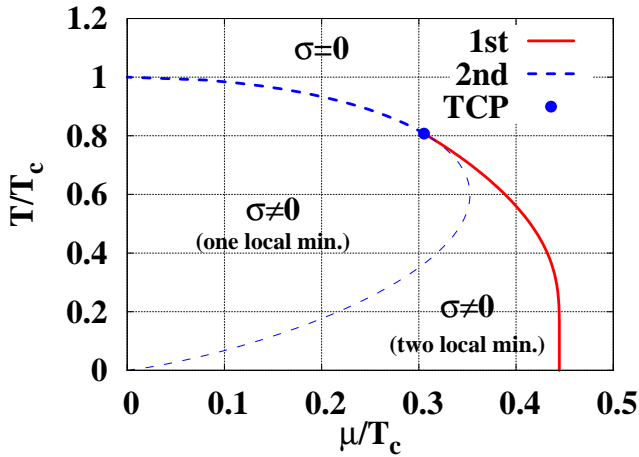}{}
\caption{
(Color online)
Phase structure in the strong coupling limit lattice QCD with $N_c=3$.
The solid (red) and upper thick dashed (blue) lines show
the first and second order phase boundary, respectively.
The dashed (blue) line shows the boundary on which 
the effective free energy curvature becomes zero at $\sigM=0$.
The dot represents the tri-critical point (TCP).
Chemical potential and temperature are shown in the unit of $T_c$.
}\label{Fig:Phase}
\end{figure}

At finite temperatures,
we can expand $\FeffQ$ in $\sigM^2$,
\beqar
&& \FeffQ(\bsig\sigM;T > 0,\mu)
\nonumber\\
&=& -T\log\left({C_{3\mu}+2\over3}\right)
	- {5\bsig^2\sigM^2\over T(C_{3\mu}+2)}
	+\Delta \FeffQ
\ ,
\eeqar
where $\Delta\FeffQ = {\cal O}(\sigM^4)$.
In the chiral limit,
the coefficient of $\sigM^2$ in $\FEFF$,
\beq
c_2(T,\mu) = \frac12\bsig- {5\bsig^2\over T(C_{3\mu}+2)}
\ ,
\eeq
controls the second order phase transition.
At zero chemical potential,
this coefficient changes sign at
\beq
T_c = T_c^\mathrm{(2nd)}(\mu=0) = {10\bsig\over 3} \ .
\eeq
For lower temperatures, $T < T_c$,
the coefficient changes sign at
a chemical potential
\beq
\label{Eq:mu_c-2nd}
\mu_c^\mathrm{(2nd)}(T) = {T\over 3}\arccosh\left(
		{3T_c\over T}-2
		\right)
\ .
\eeq
For larger chemical potentials, $\mu > \mu_c^\mathrm{(2nd)}(T)$,
the effective free energy has a local minimum at $\sigM=0$,
as already mentioned in the case of $T=0$.
The above critical chemical potential $\mu_c^\mathrm{(2nd)}$
is shown by the dashed line in Fig.~\ref{Fig:Phase}.

In the present model,
the chiral phase transition is second order
at small chemical potentials in the chiral limit.
The coefficient $c_2$ is a decreasing function of $T$ for a fixed $\mu$
for $\mu/T \lesssim 0.588$.
In addition, the higher order terms are positive when the chemical potential
is small,
\beqar
&&\Delta \FeffB + \Delta \FeffQ
\simeq c_4 \sigM^4 + {\cal O}(\sigM^6)
\ ,
\\
&&
\label{Eq:c4}
{c_4\over\bsig^4}
= {3a_0^{(b)}\over 28}\left({\gsig\Lambda\over4\bsig}\right)^4
	+{20T^2-41+150/(C_{3\mu}+2)\over 12T^3(C_{3\mu}+2)}
\ .
\nonumber\\
\eeqar
As a result, 
we do not have any local minimum at finite values of $\sigM$
giving smaller effective free energy than $\sigM=0$.
Therefore, the above $T_c$ is the actual chiral phase transition 
temperature, and then the chiral phase transition is second order
at zero and small chemical potentials in the chiral limit.
The behavior of the effective free energy at $\mu=0$ 
is shown in the upper panel of Fig.~\ref{Fig:Esurf}.

\subsection{Phase diagram}
\label{Subsec:phase}

In previous subsections, we have discussed
the properties of the effective free energy at small $\sigM$
in the chiral limit.
In order to discuss the whole phase diagram,
we show the results of numerical calculations
in the chiral limit ($m_0=0$) with a parameter value $\alpha=0.2$
in this subsection.

Since the chiral phase transition is second order
for small chemical potentials and is first order
for small temperatures,
we have to have a tri-critical point (TCP) in the phase boundary.
At TCP,
the finite equilibrium chiral condensate $\sigM$, which gives
the same effective free energy as that for $\sigM=0$,
approaches to zero.
In Fig.~\ref{Fig:Esurf},
we show the effective free energy as a function of $\sigM$
at zero chemical potential (upper panel),
zero temperature (bottom panel)
and at the TCP temperature.
We can find clear characteristic behavior
of the first order phase transition at zero temperature
and the second order phase transition at zero chemical potential.
At $T=T_\mathrm{TCP}$, we see a marginal trend.

In Fig.~\ref{Fig:Phase},
we show the phase diagram.
The dashed line shows the critical chemical potential
at which the coefficient of the quadratic term, $\sigM^2$,
becomes zero.
Outside of this dashed line, we necessarily have a local minimum
at $\sigM=0$.
At low temperatures, we have another local minimum at a finite 
value of $\sigM$, giving a lower value of the effective potential
than that of $\sigM=0$.
As a result, we have three regions in the $(T,\mu)$ plane:
The quark-gluon plasma (QGP) phase where the chiral symmetry is restored,
the region of $\mu<\mu_c^\mathrm{(2nd)}(T)$ where we have one
local minimum at a finite value of $\sigM$,
and the region $\mu_c^\mathrm{(2nd)}(T) < \mu < \mu_c^\mathrm{(1st)}$
where we have two local minima.

It is interesting to find that, with the current choice of the parameter,
$\mu_c^\mathrm{(1st)}(T)$ smoothly decreases as the temperature increases,
and it joins with $\mu_c^\mathrm{(2nd)}$ at TCP.
In the present model with one order parameter $\sigM$,
the slope of $\mu_c^\mathrm{(1st)}(T)$ (i.e. $d\mu_c^{(1st)}/dT$ ) 
in Fig.~\ref{Fig:Phase} has to be
the same as that of $\mu_c^\mathrm{(2nd)}(T)$ in the vicinity of TCP.
The first order phase transition condition of equilibrium
and balance with $\FEFF(0)$
can be solved as $4c_2 c_6 = c_4^2$
for the effective free energy
$\FEFF(\sigM) = c_0 + c_2\sigM^2+c_4\sigM^4+c_6\sigM^6$.
In the vicinity of TCP
$c_2, c_4 = {\cal O}(\Delta T, \Delta \mu)$ are small,
then the above condition requires $c_2 = {\cal O}((\Delta T,\Delta \mu)^2)$
leading to very small $c_2$ which should be on the second order phase
transition line, provided that $c_6$ is finite at around TCP.
Therefore, 
negative slope $d\mu_c^{(1st)}/dT < 0$ around TCP is a consequence
of larger TCP temperature giving a negative slope of
$\mu_c^\mathrm{(2nd)}$,
$T_\mathrm{TCP} > T_x \simeq 0.599\, T_c$.
The TCP temperature is a solution of a simultaneous equation
of $c_4=0$ and $\mu = \mu^\mathrm{(2nd)}_c(T)$,
\beq
{T_\mathrm{TCP}\over T_c}
={41\over25}\left[
	1+\sqrt{
		1+{164\over625}T_c^2
		\left(
			5 + 9T_c c_4^{(b)} / \bsig^4
		\right)
		}
	\right]^{-1}
	\ ,
\eeq
where $c_4^{(b)}$ stands for the first term of $c_4$ in Eq.~(\ref{Eq:c4}).
In the present parameterization,
the condition $T_\mathrm{TCP} > T_x$ is satisfied in a wide range,
$0.0864 \leq \alpha \leq 0.563$.
On the other hand, when we ignore the baryonic action,
we get $c_4^{(b)}=0$, $T_c=5/3$~\cite{Nishida2004b},
then $T_\mathrm{TCP} \simeq 0.52\, T_c < T_x$.

\begin{figure}
\Psfig{9cm}{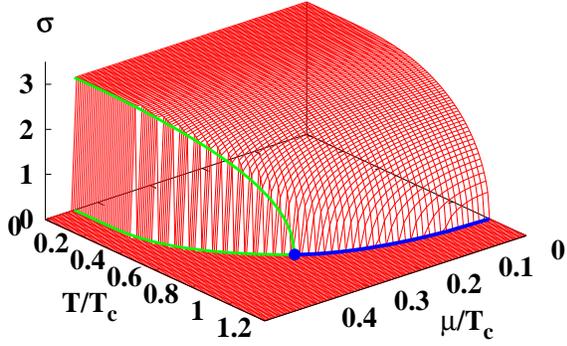}{}
\caption{
(Color online)
Chiral condensate $\sigM$ as a function
of chemical potential and temperature.
Thick lines show the phase boundary,
and the dot indicates the tri-critical point (TCP).
Chemical potential and temperature are shown in the unit
of $T_c$.
}\label{Fig:Sigma}
\end{figure}

\begin{figure}
\noindent
\Psfig{8.5cm}{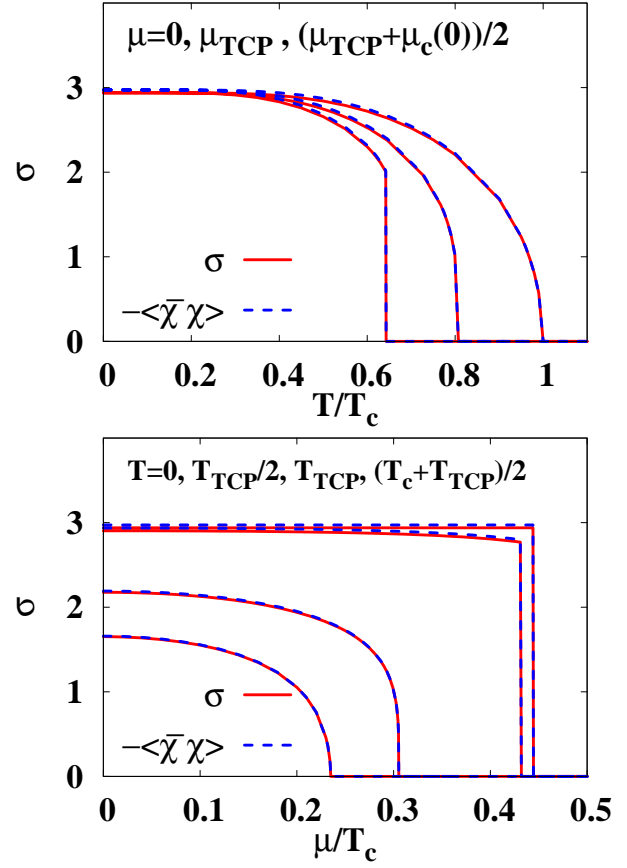}{}
\caption{
(Color online)
Approximate chiral condensate $\sigM$
and the chiral condensate $-\partial\FEFF/\partial m_0$
as a function of temperature (upper panel) and chemical potential (lower panel).
Solid (red) and dashed (blue) lines show
$\sigM$ and $-\langle\chibar\chi\rangle$,
respectively.
Chemical potential and temperature are shown in the unit of $T_c$.
}\label{Fig:SigTMu}
\end{figure}

In Figs.~\ref{Fig:SigTMu},
we show the approximate chiral condensate $\sigM$ and
the actual chiral condensate
$\langle\chibar\chi\rangle=\partial\FEFF/\partial m_0$
as a function of the temperature and chemical potential in the chiral limit.
At zero temperature, 
the effective free energy is a common function of $\sigM$
in the region $\arcsinh\,\bsig\sigM> \mu$,
and then the equilibrium value of $\sigM$
stays constant up to $\mu = \mu_c^\mathrm{(1st)}$.
As a result, the equilibrium free energy is a constant
when $\mu < \mu_c^\mathrm{(1st)}$,
and decreases as $\FEFF = \mathrm{const.}-N_c\mu$,
leading to a sudden jump of the baryon density
from $\rhoB=-\partial\FEFF/\partial(N_c\mu)=0$ to $\rhoB=1$,
which is the maximum value on the lattice.
This behavior is an artifact of the strong coupling limit,
and it is also found in previous works at zero temperature and finite chemical
potential~\cite{Bilic,BilicCleymans,Nishida2004b,ZeroT:FiniteMu}.

At finite temperatures, $\sigM$ smoothly decreases,
and suddenly vanishes at $\mu_c^\mathrm{(1st)}$ when $T<T_\mathrm{TCP}$.
The chiral condensate $\langle\chibar\chi\rangle$
is almost the same as $\sigM$.
This approximate relation holds very well
for small values of $\sigM$, 
and even for large values of $\sigM$ around the vacuum value,
the ratio changes only by around 1 \%.

\section{Extended Examinations}
\label{Sec:Discussions}

The effective free energy derived and examined in the previous sections
seems to be reasonable,
and the calculated results qualitatively agree with those in other works.
However, there are several unsatisfactory points.
First, we have to introduce two parameters, $\gamma$ and $\alpha$.
Several restrictions for these parameters are discussed in the previous
section, and further discussions is presented in Subsec.~\ref{Subsec:pardep}.
Secondly, we find quantitative differences
in some thermodynamical variables from those in other works.
In Subsec.~\ref{Subsec:Comp}, we compare the present effective free energy
and other strong coupling limit models proposed so far.
Thirdly, while we have shown that the diquark effect appears
in several aspects of the effective free energy indirectly,
it is unsatisfactory that we cannot treat the diquark 
condensate directly.
In Subsec.~\ref{Subsec:diquark}, we propose an idea how to include
the diquark condensate directly in the effective free energy.

\subsection{Parameter dependence}
\label{Subsec:pardep}

In the previous section, we have shown the relation 
between the scaled variables such as,
$T/T_c$, $\mu/T_c$ and $\FEFF/T_c$. 
This is because we can remove the major parameter dependence
with these scaled variables at small $\sigM$ values.
Here we would like to discuss that 
this scaling behavior corresponds
to the modification of the lattice spacing.

When we explicitly put the spatial and temporal lattice spacings
($a$ and $a_t$) in the effective free energy, 
we find the following dependence.
\beqar
\FEFF
&=&
	{1\over a^3 a_t}\left[
	\frac12\bsig (a^3 \sigM)^2
	+ \Delta \FeffB(a^3\gsig\sigM)
	\right]
\nonumber\\
&&~~~~~
	- {T\over a^3} \log G_U(a_t a^3\bsig\sigM;a_t T,a_t\mu)
\ ,
\label{Eq:Feff_a}
\eeqar
where $G_U = \exp(-\FeffQ/T) = \int dU_0 G(\bold{x})$.
The critical temperature depends on both of $a_t$ and $a$ as
$T_c = 10 a_t \bsig /3 a^2$.

We require that the second order chiral restoration temperature
should be described
independently from parameter choice,
when we choose the temporal and spatial lattice spacing appropriately.
Actually, the effective free energy up to $\sigM^2$,
which governs the second order chiral restoration at small $\mu$,
is found to be independent from the parameter choice for a given $T_c$,
\beqar
\FEFF
&=&
	{T_c\over a^3}\left[
	{3\sigM_a^2\over 20}
	+ \beta_c\Delta\FeffB
	\right.
\nonumber\\
&&~~
	\left.
	- {T\over T_c} \log G_U\left({3\sigM_a\over10\beta_c};{T\over\beta_cT_c},{\mu\over\beta_cT_c}\right)
	\right]
\nonumber\\
&=&
	{T_c\over a^3}\left[
		{3\sigM_a^2\over 20}
		- {T\over T_c} \log G_U
	\right]
	+ {\cal O}(\sigM^3)
\ ,
\label{Eq:Feff_2nd}
\eeqar
where $\sigM_a\equiv a^4\sigM/a_t$ stands for the chiral condensate
measured in the unit of $a_t/a^4$
and $\beta_c=1/a_t T_c$ denotes the temporal lattice size
at the critical temperature $T=T_c$.
Since $\Delta\FeffB$ starts from $\sigM^4$
and $G_U$ is a function of $\arcsinh(\bsig\sigM)/T$ and $\mu/T$,
the scaled effective free energy $\FEFF/T_c$
is a function of scaled variables
$\sigM_a$, $T/T_c$ and $\mu/T_c$
when we ignore ${\cal O}(\sigM^3)$.

\begin{figure}
\Psfig{9cm}{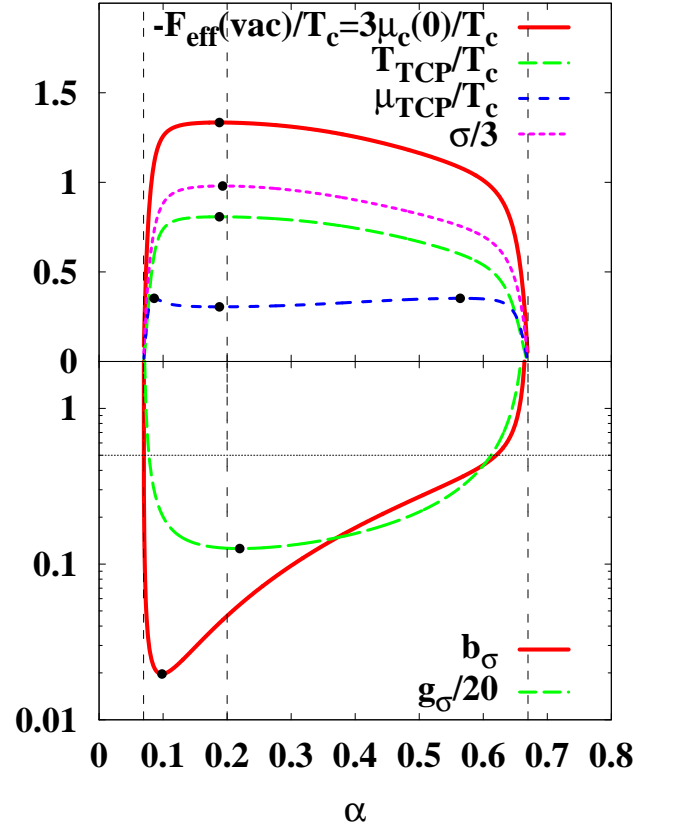}{}
\caption{
(Color online)
Parameter dependence of $\mu_c(0)$ in vacuum (thick solid line, red),
$T_{\mathrm{TCP}}$ (long dashed line, green),
$\mu_{\mathrm{TCP}}$ (short dashed line, blue),
and 
$\sigM$ (dotted line, magenta) in the upper graph.
In the lower graph, thick solid line (red) and  long dashed line (green)
represents the parameter dependence of $\bsig$ and $\gsig$,
respectively.
}\label{Fig:ParDep}
\end{figure}

The remaining parameter dependence may come from the mean field ansatz.
Thus the parameter should be chosen in the range
where the mean field ansatz is valid;
i.e. the dependence of obtained quantities is small.
In Fig.~\ref{Fig:ParDep}, we show the parameter dependence
of $\sigM$ in vacuum, $T_{\mathrm{TCP}}$, $\mu_{\mathrm{TCP}}$
and $\mu_c(0)$, which suffer from higher order contributions of $\sigM$.
Most of these quantities have extrema at around $\alpha=0.2$
($\alpha \simeq 0.188$ for
$T_{\mathrm{TCP}}$,
$\mu_{\mathrm{TCP}}$
and $mu_c(0)$,
and $\alpha \simeq 0.193$ for $\sigM$ in vacuum).
In addition, we find that the parameter dependence is not strong
in the parameter range, $0.1 \leq \alpha \leq 0.6$.
It is worth to mention here that $\bsig$ is small enough
at around $\alpha \simeq 0.2$
and it satisfies the even integer condition for
$\beta_c = 1/a T_c = 3 / 10 \bsig \geq 2$
in a symmetric lattice.

\begin{figure}
\Psfig{9cm}{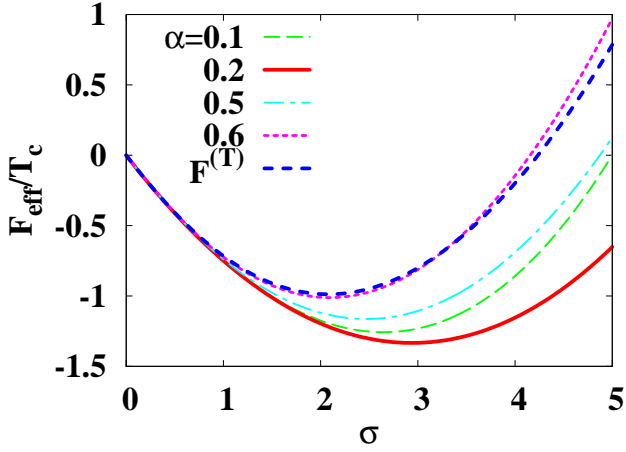}{}
\caption{
(Color online)
Parameter dependence of the effective free energy 
in vacuum in the chiral limit, $(T,\mu,m_0)=(0,0,0)$.
Long dashed (green), thick solid (red), dash-dotted (light blue),
and dotted (magenta) lines show the effective free energy
in the present work, $\FEFF^{(Tb)}$, with parameters
$\alpha=$ 0.1, 0.2, 0.5 and 0.6, respectively.
Short dashed line (blue) shows the effective free energy
without baryon effects, $\FEFF^{(T)}$.
}\label{Fig:Surfpar}
\end{figure}

\begin{figure}
\Psfig{9cm}{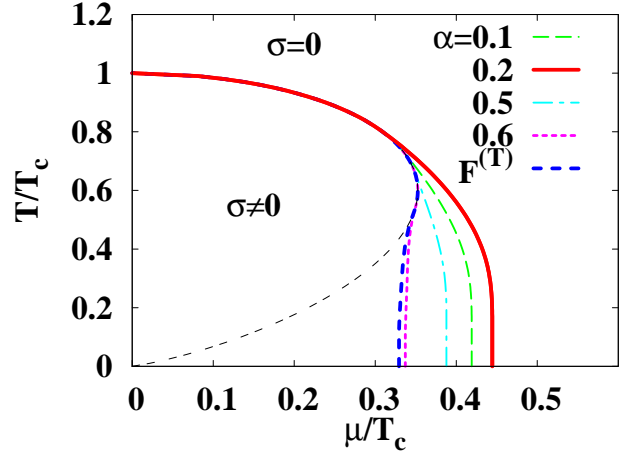}{}
\caption{
(Color online)
Parameter dependence of the phase boundary.
Long dashed (green), thick solid (red), dash-dotted (light-blue),
and dotted (magenta) lines show the phase boundary in $\FEFF$
with $\alpha=$0.1, 0.2, 0.5 and 0.6, respectively.
Short dashed line (blue) shows the boundary in $\FEFF^{(T)}$,
without baryon effects.
Thin solid line (black) indicates $c_2=0$,
where $c_2$ is the quadratic coefficient of $\sigM$.
}\label{Fig:Phase-Comp}
\end{figure}

In Figs.~\ref{Fig:Surfpar} and \ref{Fig:Phase-Comp},
we show the parameter dependence
of the effective free energy in vacuum and the phase boundary.
In these figures, we also plot the results
with the effective free energy $\FEFF^{(T)}$, 
\begin{equation}
\label{Eq:FEFF-T}
\FEFF^{(T)}=\frac12\bsig^{(T)}\sigM^2
	+\FeffQ(\mq)
=T_c \left[
	{3\over 20}\sigM^2
	- {T\over T_c} \log G_U
	\right]
\ ,
\\
\end{equation}
\begin{equation}
\mq=\bsig^{(T)}\sigM+m_0 ,\ 
\bsig^{(T)}={d\over2N_c},\ 
T_c={10\bsig^{(T)}\over3}=\frac53
\ ,
\end{equation}
which is obtained by ignoring
the baryon effects and integrating over $U_0$
exactly in a similar way to that
in Ref.~\cite{Bilic,BilicCleymans,Nishida2004b}.
It is clearly seen that large energy gain is obtained
with $\alpha \simeq 0.2$,
and the phase boundary extends to the larger $\mu$ direction.
When we ignore baryon effects,
the effective free energy and phase boundary with $\FEFF^{(T)}$
roughly corresponds to those with $\alpha=0.6$
in the present model with baryons.

\subsection{Comparison with other treatments}
\label{Subsec:Comp}

While we have treated the time-like link variable $U_0$ exactly
in the previous section,
the anti-periodic boundary condition may not be very important
when the temporary lattice size, $\beta = 1/T$, is very large.
In this case, it is possible to perform the one link integral
also for $S_F^{(U_0)}$ as other spatial action, $S_F^{(U_j)} (j=1,2,3)$.
After introducing auxiliary fields, $b, \barb$ and $\sigM$,
we obtain the action,
\beqar
&&\int \Fint{U} e^{-S_F[U,\chi,\chibar]}
 \simeq \int \Fint{b,\barb,\sigM} e^{-S_F^{(0)}-S_F^{(m)}}
\ ,\\
&&S_F^{(0)}[\chi,\chibar,b,\barb,\sigM]
=\frac12\sum_{x,y}\sigM(x)V_{M0}(x,y)\sigM(y)
\nonumber\\
&&+\sum_{x,y}\left[\barb(x)V_{B\mu}^{-1}(x,y) b(y)
+\sigM(x) V_{M0}(x,y) M(y)\right]
\nonumber\\
&&-\sum_x\left[\barb(x){B}(x)+\bar{B}(x)b(x)\right]
\ ,\\
&&
V_{M0}(x,y)={1\over 4N_c}
\sum_{\nu=0}^3
\left(\delta_{y,x+\hat{\nu}} + \delta_{y,x-\hat{\nu}}\right)
\ ,\\
&&
V_{B\mu}(x,y)=
V_{B}
-\frac18
\left(e^{3\mu}\delta_{y,x+\hat{0}}-e^{-3\mu}\delta_{y,x-\hat{0}}\right)
\ .
\eeqar

\begin{figure}
\Psfig{9cm}{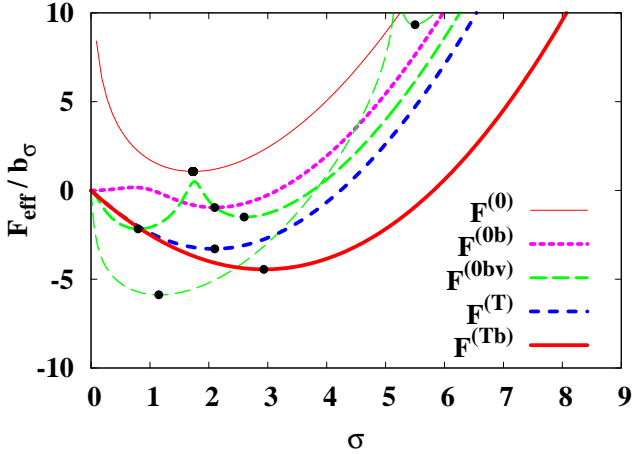}{}
\caption{
(Color online)
Comparison of effective free energies in different treatments.
Thin solid (red),
dotted (magenta),
long-dashed (green), 
and short-dashed (blue)
lines show the effective free energies
$\FEFF^{(0)}$, $\FEFF^{(0b)}$, $\FEFF^{(0bv)}$  and $\FEFF^{(T)}$,
respectively.
For $\FEFF^{(0bv)}$ (long-dashed, green),
we show the results with two parameters;
$\alpha=\sqrt{1/2-\gamma^2}=0.2$ (thick) and $\gamma=2$ (thin).
The thick solid line (red) indicates the effective free energy
in a finite temperature treatment with baryon effects
($\FEFF^{(Tb)}$, present work).
We show the results in vacuum in the chiral limit,
$(T,\mu,m_0)=(0,0,0)$.
}\label{Fig:Model-Comp}
\end{figure}

In the above action,
quark fields completely decouples in each space-time point,
then it is possible to perform quark integral.
For example, when we ignore the baryon effects
and carry out the quark integral,
we obtain the following effective free energy \cite{KS1981}, 
\beqar
\FEFF^{(0)}
&=& \frac12\bsig^{(0)}\sigM^2-N_c\log(\bsig^{(0)}\sigM+m_0)
\ ,
\eeqar
where $\bsig^{(0)}=(d+1)/2N_c$ and $\mq=\bsig^{(0)}\sigM+m_0$.
The diverging behavior at $\sigM=0$ in the chiral limit is suppressed
when we include the baryon effects in a similar way to that in ~\cite{DHK1985},
\beqar
\label{Eq:Feff0b}
\FEFF^\mathrm{(0b)}
&=& \frac12\bsig^{(0)}\sigM^2+\Feff^{(b\mu)}(4\mq^3;T,\mu)
\ .
\eeqar
The expression of the baryon integral, $\Feff^{(b\mu)}$ is shown
in the Appendix \ref{App:Baryon}.
It is also possible to obtain the effective action
with diquark field as,
\beqar
\label{Eq:Feff0bv}
\FEFF^\mathrm{(0bv)}
&=&\frac12\bsig^{(0)}\sigM^2
+ v^2 - \log\Theta + \Feff^{(b\mu)}(m;T,\mu)
\ ,
\\
\Theta &=& \frac13\left(\frac{1}{R_v^2} - {\mq^2\over R_v\gamma^2}+\frac29 v^2\right)
\ ,
\\
m &=& {4\mq\left(3\gamma^2/R_v-\mq^2\right)\over\Theta}
\ ,
\eeqar
where $R_v\equiv 1-v^2/3$.
This effective free energy is essentially the same as that
in Ref.~~\cite{Azcoiti},
while we use a different notation
and introduce a parameter $\gamma$ as in the previous section.

These effective free energies have similar asymptotic behavior
for large $\sigM$ in vacuum.
Knowing the asymptotic form,
$\Feff^{(b\mu)}(m)\to-\log 2m$ at $m \to \infty$,
we find all the potential terms
in $\FEFF^{(0)}, \FEFF^{(0b)}$ and $\FEFF^{(0bv)}$,
have the form of $-N_c\log\sigM+ \mathrm{const.}$
in the large $\sigM$ limit.
The effective free energy at finite $T$,
$\FEFF^{(T)}$ in Eq.~(\ref{Eq:FEFF-T}),
also has the potential of the above form,
since $\FeffQ(\sigM)\to-\log2\sigM(\sigM\to \infty)$.

In Fig.~\ref{Fig:Model-Comp},
we compare the effective free energies
as a function of the chiral condensate $\sigM$
in vacuum in the chiral limit.
We show the scaled effective free energies
$\FEFF^{(i)}/\bsig^{(i)}$ instead of $\FEFF^{(i)}/T_c$,
since the chiral restoration does not emerge 
with zero temperature effective free energies,
$\FEFF^{(0)}, \FEFF^{(0b)}, \FEFF^{(0bv)}$.
In $\FEFF^{(0bv)}$,
the diquark condensate $v$ is set to be zero,
as the global minimum is already reported to lie at $v=0$~\cite{Azcoiti},
and the results with two parameters are compared;
the same parameter set as in the present work,
$\gamma=\sqrt{1-\alpha^2}$ with $\alpha=0.2$,
and the value originally adopted in Ref.~\cite{Azcoiti}, $\gamma=2$.
From a comparison of $\FEFF^{(0)}$ and $\FEFF^{(0b)}$, 
the main role of baryons is found
to reduce the effective free energy at small $\sigM$ values,
in addition to suppressing the diverging behavior at $\sigM=0$.
On the other hand, in zero temperature treatments with baryons,
$\FEFF^{(0b)}$ and $\FEFF^{(0bv)}$,
we find a bump at a small $\lambda$, which separates two local minima.
This bump comes from the slow startup of the baryon contribution
proportional to $\sigM^6$ at small $\sigM$ in $\FEFF^{(0b)}$,
and from a cancellation between $-\log\Theta$ and $\Feff^{(b\mu)}(m)$
in $\FEFF^{(0bv)}$.
In a finite temperature treatment, $\FEFF^{(T)}$,
reduction effect is smooth, and we find only one local minimum.
The effective free energy in the present work,
$\FEFF^{(Tb)} = \FEFF$ in Eq.~(\ref{Eq:Feff}), 
is smaller than those in other treatments
except for $\FEFF^{(0bv)}$ with $\gamma=2$, 
with which the effective free energy becomes unstable
in a finite temperature treatment.
The large energy gain in $\FEFF^{(Tb)}$ may partly come from 
the scaling of $\FEFF/\bsig$, since the $\bsig$ in this work
is the smallest among the models compared here.
In order to compare the absolute values of the effective free energy
more seriously,
it would be necessary to fix the lattice spacing
and thus the energy scale
and to include the effects of the higher order contribution
in the $1/d$ expansion.

\subsection{Expected evolution of the phase diagram}

One of the common problems in the strong coupling limit
is the too small critical chemical potential
$\mu_c(T=0)$ relative to $T_c=T_c(\mu=0)$.
In Table~\ref{Table:Ratio},
we compare the ratio of
the critical baryon chemical potential at zero temperature
with respect to the critical temperature at zero chemical potential,
$R_{\mu T} \equiv 3\mu_c(T=0)/T_c$.
All the models based on the strong coupling limit 
give much smaller values for this ratio, $R_{\mu T} < 1.5$,
than the empirical value,
$R_{\mu T} =(1-2)\mathrm{GeV}/170 \mathrm{MeV} \simeq (6-12)$.
In the Monte-Carlo simulations at finite quark chemical potentials
with finite $1/g^2$,
it is not yet possible to obtain $\mu_c(T=0)$,
but larger $R_{\mu T}$ values are suggested.
For example,
several Monte-Carlo methods are in agreement with each other
for small quark chemical potentials
$\mu/T < 1$~\cite{LatticeFiniteMu,Taylor,AC,Canonical,FodorKatz},
and the critical temperature for these chemical potentials are large enough,
$T_c(\mu)/T_c(\mu=0) \gtrsim 0.9$,
implying that $R_{\mu T} \gg 3$.

\begin{table}
\caption{
Ratio of the critical chemical potential at zero temperature
$\mu_c(T=0)$ and the critical temperature $T_c=T_c(\mu=0)$
in strong coupling models.
In $\FEFF^{(0b)}$, we have assumed $T_c=5/3$ to obtain the ratio.
For the results in Ref.~\protect\cite{Bilic},
values of critical lattice anisotropy $a/a_t$
and that multiplied by $\mu$ are taken
for the number of staggered fermions, $f=1,2$ and 3.
}\label{Table:Ratio}
\begin{tabular}{l|cccl}
\hline
\hline
Model		& $T_c$	& $\mu_c(T=0)$	& $3\mu_c(0)/T_c$ &\\
\hline
$\FEFF^{(0b)}$~\protect\cite{DHK1985}
	&     & 0.66	& 1.19	& ($T_c=5/3$)	  \\
$\FEFF^{(T)}$~\protect\cite{Nishida2004b}
	& $5/3$		& 0.33 $T_c$	& 0.99		\\
$\FEFF^{(Tb)}$
	& $10\bsig/3$	& 0.45 $T_c$	& 1.34  & ($\alpha=0.2$)\\
\hline
Ref.~\protect\cite{Bilic}
	& 2.57 & 0.57 & 0.67 & ($f=1$) \\
	& 2.19 & 0.57 & 0.78 & ($f=2$) \\
	& 2.07 & 0.57 & 0.83 & ($f=3$) \\
\hline
Empirical & 170 MeV	& (1-2)/3 GeV & $\sim$ (6-12) \\
\hline
\hline
\end{tabular}
\end{table}

\begin{figure}
\Psfig{6.5cm}{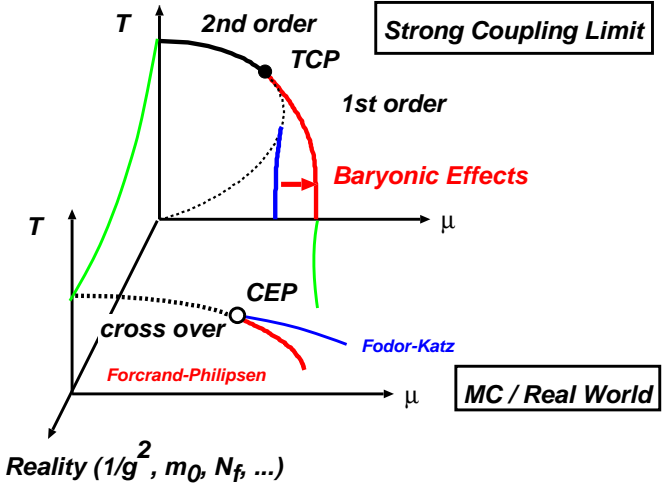}{,angle=-90}
\caption{
(Color online)
Expected phase diagram evolution
from the strong coupling limit in the chiral limit with one species of 
staggered fermion towards the real world.
}\label{Fig:Phase-Evol}
\end{figure}

Thus for a quantitative discussion,
the strong coupling limit in the chiral limit with one species of staggered
fermion is not enough,
and it is necessary to take care of
finite quark mass $m_0$,
multi-staggered fermions,
finite $1/g^2$,
other order parameters than the chiral condensate,
and/or other mechanisms towards the real world
in order to explain large $\mu_c(T=0)$ relative to $T_c$,
as illustrated in Fig.~\ref{Fig:Phase-Evol}.
With finite quark mass $m_0$,
the effective free energy $\FEFF$ always has a minimum at finite $\sigM$,
then the second order boundary becomes cross over
and the tri-critical point (TCP) becomes the critical end point (CEP).
In addition,
since the finite quark mass $m_0$ increases the baryon mass
which is closely related to $\mu_c(T=0)$~\cite{DHK1985},
finite $m_0$ is believed to increase $\mu_c(T=0)$,
as shown for example in Ref.~\cite{Nishida2004b}.
With multi-staggered fermions,
$T_c$ is suppressed as discussed in Ref.~\cite{Bilic}.
It would be natural to expect that $T_c$ decreases as $1/g^2$ grows,
because hadrons and glueballs are more bound at larger couplings
and thus hadronic phase would be the most stable in the strong coupling limit.
We further expect that the finite coupling effects appears most strongly
at $\mu=0$, where the role of gluons relative to quarks is the largest.
Actually in Ref.~\cite{BilicCleymans},
it is shown that $T_c(\mu)$ decreases as $1/g^2$ increases,
and this reduction is more rapid at $\mu=0$ than at finite $\mu$.

With other order parameters than the chiral condensate,
the phase diagram will have a richer structure.
In Subsec.~\ref{Subsec:phase}, we have discussed that
the slope of $\mu_c^{\mathrm{(1st)}}$ (see Fig. \ref{Fig:Phase})
have to be the same
as that of $\mu_c^{\mathrm{(2nd)}}$ at the vicinity of TCP
in the chiral limit with one order parameter.
For this point, there is a debate between two Monte-Carlo simulations,
one of which suggests the smooth connection of cross over boundary
and the first order boundary~\cite{Taylor,AC,Canonical},
and the other suggests a finite difference in slope~\cite{FodorKatz}.
Provided that the nature of TCP remains in CEP
even with finite quark mass,
our discussion in Subsec.~\ref{Subsec:phase} supports the former
if there is only one order parameter.
However, both of the above two results can be consistent
if there are other order parameters than the chiral condensate.
Smooth connection is expected for chiral transition
in methods based on the analyticity~\cite{Taylor,AC,Canonical},
while we may see other transition
in a direct Monte-Carlo method at finite chemical potentials~\cite{FodorKatz}.

\subsection{Color angle average in diquark condensate}
\label{Subsec:diquark}

In deriving the effective free energy in Eq.~(\ref{Eq:Feff})
we have assumed that the diquark condensate takes zero values, $\phi_a=0$.
If we ignore the diquark-gluon-baryon coupling,
$S_F^{(C)}$ in Eq.~(\ref{Eq:SFC}),
it is possible to obtain the solution of $\det \bold{g}_{ab}(k_m)=0$
and the integral over $U_0$ numerically.
With $S_F^{(C)}$, however,
since $U_0$ depends on $\bold{x}$
and baryon fields are spatially connected through $\VB$,
we have to carry out the integral of baryon determinants
over $2L^3$ dimensional variables in $U_0(\bold{x})$.
In this subsection, we would like to show an idea to solve this problem
in a case with one species of staggered fermion.

Since the diquark field $\phi_a$ is not color singlet,
its average over the color space should be zero.
Thus we cannot treat it as an order parameter.
One of the way to remedy this problem is to carry out
the integral of the "color angle" variables in $\phi_a$,
then only the color singlet combination $v^2=\phi_a^\dagger\phi_a$
remains.
It is possible to carry out this color angle average in a straightforward way.
\beqar
&&
	\Bigl\langle
	\exp\left[
		\sum_a\left(
		\phi_a^\dagger D_a+D_a^\dagger \phi_a
		\right)
	\right]
	\Bigr\rangle_v
\nonumber\\
&=&
	\Bigl\langle
	\prod_a \left[
		1 + \phi_a^\dagger\phi_a D_a^\dagger D_a
		+\frac14 (\phi_a^\dagger\phi_a)^2 (D_a^\dagger D_a)^2
		\right]
	\Bigr\rangle_v
\nonumber\\
&=&
	1 + \sum_a \langle\phi_a^\dagger\phi_a\rangle_v D_a^\dagger D_a
	+\frac14 \sum_a \langle(\phi_a^\dagger\phi_a)^2\rangle_v
		(D_a^\dagger D_a)^2
\nonumber\\
&&~~~+	\sum_{a<b}
	\langle(\phi_a^\dagger\phi_a)(\phi_b^\dagger\phi_b)\rangle_v
		(D_a^\dagger D_a) (D_b^\dagger D_b)
\nonumber\\
&=&
	1 + {v^2\over3} \sum_a D_a^\dagger D_a
	-{M^3\barb{b}\over54}
		\Bigl\langle
		\left(\sum_a \phi_a^\dagger\phi_a\right)^2
		\Bigr\rangle_v
\nonumber\\
&=&
	\exp\left(
		{v^2\over3} \sum_a D_a^\dagger D_a
		+{v^4\over 162} M^3\barb{b}
	\right)
\ .
\eeqar
Here we explicitly show the sum or product over color indices,
and $\langle\cdots\rangle_v$ means the color angle average.
When we integrate over the phase variable for each $\phi_a$,
we only have those terms having the same power of $\phi_a$ and $\phi_a^\dagger$
as shown in the second line.
The power of the diquark composite $D_a$ and $D^\dagger_a$ is limited
to four as shown in the third line by the Grassmann nature.
For example, the power four terms such as
\beqar
\frac12 (D_1^\dagger D_1)^2 =
(D_1^\dagger D_1)(D_2^\dagger D_2)
	&=& -{1\over27}M^3\barb{b}
\ ,
\label{Eq:D12}
\eeqar
already contain all the Grassmann variables,
and product with other $D_a$ or $D^\dagger_a$ vanishes.
By using Eq.~(\ref{Eq:D12}),
we find that the fourth order terms in $D_a$ and $D_a^\dagger$
can be arranged in the form of $(\sum\phi_a^\dagger\phi_a)^2=(v^2)^2$.
In the second order terms,
we use the symmetry for each color index,
for example, $\langle\phi_1^\dagger\phi_1\rangle_v=v^2/3$.

Unfortunately we again have the term containing the coupling
of three-quark and baryon, $\barb{B}+\bar{B}b$,
from 
$D_a^\dagger D_a = Y + \barb{B}+\bar{B}b$ (see Eq.~(\ref{Eq:DDY})).
\beqar
&&\exp(\barb{B}+\bar{B}b)
\nonumber\\
&=&\int \Fint{v}
 e^{-v^2-Y}
	\Bigl\langle
	e^{\phi^\dagger D+D^\dagger \phi}
	\Bigr\rangle_v
\nonumber\\
&=&\int \Fint{v}
 e^{
 -v^2-Y+v^2\left(\barb{B}+\bar{B}b+Y\right)/3
	+v^4M^3\barb{b}/162
 }
\nonumber\\
&\propto&
 e^{
 v^2\left(\barb{B}+\bar{B}b\right)/3
 	-v^2-R_v Y +v^4M^3\barb{b}/162
 }
\ ,\\
&&
R_v = 1-v^2/3
\ .
\eeqar
In the third line, we have assumed that
the integral in r.h.s. can be approximated by the representative value of $v$.
This approximation would be valid when the diquark condensate is strong.
In this mean field ansatz, then
we can solve this self-consistent relation as follows,
\beqar
e^{R_v(\barb{B}+\bar{B}b)}
&\simeq&
 e^{-v^2-R_vY +v^4M^3\barb{b}/162}
\ ,
\eeqar
where we have ignored the constant shift in the exponent.
As a result, we obtain the following relation,
\beqar
\exp(\barb{B}+\bar{B}b)
&\simeq&
	\exp\left[ -{v^2\over R_v}+{v^4 M^3\barb{b}\over162R_v} -Y\right]
\ .
\label{Eq:SCC}
\eeqar
Coupling terms of $M^n \barb{b}$ can be bosonized
by introducing $n$ bosons, whose expectation values are related
to the product of $\bar{q}q$ pair and $\barb{b}$ pair $M^k\barb{b}$,
$\VEV{\omega_k}=-\alpha_k\VEV{M}+\beta_k\VEV{M^k\barb{b}}$,
in a similar way to that in Sec.~\ref{Sec:Model}.
After introducing three auxiliary fields, $\omega_2, \omega_1, \omega_0$,
it is possible to carry out the Grassmann variable integral,
$b$ and $\chi$, and we obtain the effective free energy.
\beqar
\label{Eq:FeffV}
\FEFF^{(Tbv)}
	&=& F_X(\sigM,v,\omega_i)
		+\FeffB(g_\omega\omega)
		+\FeffQ(\mq)
\ ,\\
F_X
	&=&\frac12 (\asig \sigM^2+\omega^2+\omega_1^2+\omega_2^2)
	+ {v^2\over R_v}
\ ,\\
\asig&=&\frac12-\gamma^2-\alpha^2-\alpha_1^2-\alpha_2^2
\ ,\\
\mq &=& \asig\sigM+ \alpha\omega + \alpha_1\omega_1 + \alpha_2\omega_2
	+m_0
\ ,\\
g_\omega &=& {1\over 9\alpha\gamma^2}
		\left[
		1+{\gamma^2v^4\omega_1\omega_2\over18\alpha_1\alpha_2R_v}
		\right]
\ .
\eeqar
Here we have replaced $\omega_0=\omega$ and $\alpha_0=\alpha$.

With zero diquark condensate,
the effective free energy $\FEFF^{(Tbv)}$ in Eq.~(\ref{Eq:FeffV}),
has a similar structure to $\FEFF$ in Eq.~(\ref{Eq:Feff}).
Specifically, when we take the linear approximation
in the same way to that in Eq.~(\ref{Eq:Linear}),
we find that the chiral condensate polarizability
and the coupling constant are the same as before,
$\bsig^{(Tbv)}=\bsig, \gsig^{(Tbv)}=\gsig$
defined in Eq.~(\ref{Eq:bsiggsig}),
and we obtain the same effective free energy as before
defined in Eq.~(\ref{Eq:Feff}),
\beqar
&&\FEFF^{(Tbv)}(v=0)
\nonumber\\
&&\simeq \frac12\bsig^{(Tbv)}\sigM^2
	+ \FeffQ(\bsig^{(Tbv)}\sigM+m_0)
	+ \Delta\FeffB(\gsig^{(Tbv)}\sigM)
\nonumber\\
&&~~=\FEFF
\ .
\eeqar
This equivalence may serve a cross check of the effective free energy
derived in Sec.~\ref{Sec:Model}.

On the other hand,
there is no potential effects proportional to $v^2$
while we have quadratic term in $F_X$,
then we will not have any second order phase transition 
to the diquark condensed state.
This comes from the cancellation in Eq.~(\ref{Eq:SCC})
in the case of one staggered fermion.
Detailed analysis of the effective free energy Eq.~(\ref{Eq:FeffV})
and its extension with multi-staggered fermions~\cite{DKS1986,Bilic}
will be reported elsewhere.

\section{Summary}
\label{Sec:Summary}

In this work, 
we have studied the phase diagram of QCD for color SU(3)
at finite temperature ($T$) and finite chemical potential ($\mu$)
by using an effective free energy derived in the strong coupling limit
including baryon effects.
We have adopted the effective action
up to the next-to-leading order of the $1/d$ expansion
(${\cal O}(1)$ and ${\cal O}(1/\sqrt{d})$),
and by using the mean field ansatz,
an analytical expression of the effective free energy is derived.
The baryonic composite term in the effective action is decomposed
into the terms consisting diquark condensates, baryons,
and quarks~\cite{Azcoiti}.
By introducing auxiliary fields of the baryon, diquark, baryon potential,
and chiral condensate, we have obtained the effective action
in the bilinear form of fermions.
Then the Grassmann integral of quarks and the sum over the Matsubara
frequency can be carried out exactly,
provided that the solution of $\det \bold{g}_{ab}(k_m)=0$ is obtained.
At zero diquark condensate, 
we can further perform the integral
over the temporal link variables and baryon fields analytically.

This is the first trial which introduces baryon and finite temperature
effects simultaneously in the strong coupling limit of lattice QCD
for color SU(3).
It is important to note that baryon has effects
to reduce the effective free energy $\FEFF$ as shown in Fig.~\ref{Fig:Surfpar}
and to extend the hadron phase to a larger $\mu$ direction
at low temperatures as shown in Fig.~\ref{Fig:Phase-Comp},
when $\FEFF$, $\mu$ and $T$ are measured relative to $T_c$.
We may expect that this feature remains in the realistic parameter
region of finite $1/g^2$.
It would then be interesting to compare the phase boundary behavior
between SU(3) and U(3) to examine the baryon effects
in this parameter region.

The obtained phase diagram have
the second order phase boundary at small chemical potentials, 
and the first order phase boundary at small temperatures
separated by a tri-critical point.
This feature is the same as that in previous works,
but the ratio of the critical baryon chemical potential at zero temperature
with respect to the critical temperature at zero chemical potential,
$R_{\mu T} \equiv 3\mu_c(T=0)/T_c(\mu=0)$,
is found to be much smaller
than the empirical value or that suggested in Monte-Carlo simulations.
Small $R_{\mu T}$ is a common feature
in models based on the strong coupling limit,
and it would be necessary to extend in the direction of the reality axes
in Fig.~\ref{Fig:Phase-Evol} for a quantitative discussions.
On the other hand, we expect that
Monte-Carlo simulations should reproduce
the strong coupling results
of the phase boundary
including the small value of $R_{\mu T}$
at a large value of $g$.

Finally, we have proposed a method,
color angle average in colored auxiliary fields, 
which enables us to extract a color singlet order parameter
and to include the diquark condensate explicitly in the effective free energy.

One of the problems which we have found in this work 
is the parameter dependence of the effective free energy $\FEFF$.
During the bosonization, we have introduced two parameters,
$\alpha$ and $\gamma$.
Since these are introduced through identities, 
the results should not strongly depend on the parameter choice
and in fact we have shown that 
we can absorb a major parameter dependence
of $\FEFF$ at small $\sigM$ values,
which determines the second order chiral transition,
in the choice of the lattice spacing.
For the remaining parameter dependence,
we have required that the scaled effective free energy $\FEFF/T_c$
in vacuum becomes as small as possible,
and we have adopted $\alpha=0.2$ and $\alpha^2+\gamma^2 = 1/2 - 0$.
This choice of parameters results in the temporal lattice spacing
of $\beta_c = 1/a T_c(\mu=0) \simeq 6.45$ in a symmetric lattice.
However, we have learned that $1/a T_c(\mu=0)$ is not large and less than two
in the strong coupling Monte-Carlo simulations
with one species of staggered fermion (without quarter root
of the quark determinant as in the present work)~\cite{ForcrandPrivate}.
It means that the parameter region, $0.5 \leq \alpha \leq 0.6$, is preferred,
rather than $\alpha \simeq 0.2$.
Considering these situations, we have to agree that
the baryonic effect on the phase diagram delicately depends
on the choice of the parameter $\alpha$,
and it is desired to find a general procedure to determine them.

There are several future issues to be studied further:
The first one is an extensive analysis
of the effective free energy at finite quark masses
and/or with diquark condensates.
Secondly, interesting and promising direction
is to consider multi species of staggered fermions,
since color superconductor is expected to emerge at high densities
when multi quark flavors are introduced~\cite{CSC}.

\section*{Acknowledgements}
The authors would like to acknowledge
Yusuke Nishida
and Philippe de Forcrand
for fruitful discussions and useful suggestions.
This work is supported in part by the Ministry of Education,
Science, Sports and Culture,
Grant-in-Aid for Scientific Research
under the grant numbers,
    13135201		
and 15540243.		

\appendix

\section{Sum over the Matsubara frequencies}
\label{App:Sum}
The quark determinant $G(\bold{x})$
appeared in Eq.~(\ref{Eq:detg}) is an even function of $k_m$,
then it may be expressed as a polynomial of $\cos{k_m}$.
\begin{eqnarray*}
G(\bold{x})&\equiv&\prod_{m=1}^{\beta/2}
\mbox{det}\left[\bold{g}_{ab}(k_m)\right]
=\prod_{m=1}^{\beta}
\mbox{det}\left[\bold{g}_{ab}(k_m)\right]^{1/2}\\
&=&\prod_{m=1}^{\beta}\prod_{j=1}^6
\left(\cos{k_m} - r Y_j(\bold{x})\right)^{1/2}
\end{eqnarray*}
where $r=1$.
The derivative by $r$ reads
\beqar
{d\log{G}(\bold{x}) \over dr}
&=& \frac12\sum_{m,j} {-Y_j \over \cos{k_m} - r Y_j}
\nonumber\\
&=&{1\over 2\Omega} \sum_j \left[
		\oint {dz \over 2\pi i} {-Y_j \over \cos{z}-rY_j}
			{-i\beta \over 1 + e^{i\beta z}}
	\right.
\nonumber\\
&&~~~~
	\left.
	- \sum_{z_j^r} {-Y_j\over - \sin{z_j^r}}
			{-i\beta \over 1 + e^{i\beta z_j^r}}
	\right]
\nonumber\\
&=&{i\beta\over 2\Omega} \sum_{j,z_j^r}
	{Y_j\over \sin{z_j^r}}
			{1 \over 1 + \exp(i\beta z_j^r)}
\nonumber\\
&=&{-i\beta\over 2\Omega} \sum_{j,z_j^r}
	{dz_j^r\over dr}
			{1 \over 1 + \exp(i\beta z_j^r)}
\nonumber\\
&=&{d\over dr} {1\over 2\Omega} \sum_{j,z_j^r}
		\log\left( 1 + \exp(-i\beta z_j^r)\right)\label{del_log_G}
\ .
\eeqar
In the contour integral,
we have contributions from $z = k_m$ poles
as well as $z = z_j^r$, whose sum becomes zero.
Here $z_j^r$ is the solution of $\cos{z} = rY_j$,
and $\Omega$ stands for the degeneracy for $z_j^r(+ 2n\pi i)$.
In the second line from the bottom in Eq. (\ref{del_log_G}), 
we have used the relation,
\beq
{d\cos{z_j^r}\over dr} = -\sin{z_j^r} {dz_j^r \over dr}=Y_j
\ .
\eeq

Now we obtain $\log{G}$ up to a factor.
\beq
\log{G}(\bold{x}) = {1\over 2}
	\sum_{j}\sum_{z_j} \log\left(1 + \exp(-i\beta z_j) \right)
	+ \mbox{const.}
\ .
\eeq
The sum over $z_j$ is understood as we ignore the degeneracy $2n\pi i$,
but we still have two solutions in a pairwise way, $\pm z_j$,
since $z_j$ is the solution of $\cos{z}=Y_j$.
We choose one of them as a principle value.
\beqar
\log{G}(\bold{x})&=&{1\over 2}\sum_{j}
		 \log\left[
		 	(1+e^{-i\beta z_j})
			(1+e^{i\beta z_j})
		\right]
	+ \mathrm{const.}
\nonumber\\
	&=& 
	\frac12
	\sum_{j}
		 \log(1+\cos\beta z_j)
	+ {\rm const.}
	\ .
\eeqar
We ignore the constant terms in $\log{G}(\bold{x})$,
and we get $G(\bold{x})$ as follows.
\beq
G(\bold{x}) = \left[ \prod_{j}
		 \left( 1 + \cos\beta z_j(\bold{x}) \right)
		\right]^{1/2}
\ .
\eeq

\section{Baryon Integral}
\label{App:Baryon}

In this appendix, we show how to obtain the baryon determinant
$\mathrm{Det}(1+\omega V_B)$.

First, we make a Fourier transformation of baryon field,
\beqar
b_m(\bold{x})
= {1\over\sqrt{L^3}}\sum_{\bold{k}} e^{i\bold{k}\cdot\bold{x}} b_{m\bold{k}}
\ ,\quad
\bold{k} = {2\pi\over L} (k_1, k_2, k_3)
\ .
\eeqar
The staggered factor $\eta_j(x)$
in $V_B$ connects four different momenta,
\beqar
\bold{k}^{(1)}=(k_1,k_2,k_3) &,&\quad
\bold{k}^{(2)}=(k_1+\pi,k_2,k_3)\ ,
\nonumber\\
\bold{k}^{(3)}=(k_1+\pi,k_2+\pi,k_3) &,&\quad
\bold{k}^{(4)}=(k_1,k_2+\pi,k_3)\ ,
\nonumber\\
\eeqar
and two different frequencies, $m$ and $m+\beta/2$.
As a result, $V_B$ is found to be block diagonal.
\beqar
&&
\sum_{m,n,\bold{k},\bold{k}'}
	\left(\bar{b}_{m\bold{k}},V_B b_{n\bold{k}'}\right)
\nonumber\\
&=&
\sum_{k_1,k_2=1}^{L/2}
\sum_{k_3}^L
\sum_{m=1}^{\beta/2}
\begin{pmatrix}
\bar{\bold{b}}_m & \bar{\bold{b}}_m'\\
\end{pmatrix}
\begin{pmatrix}
0 & \bold{S}(\bold{k}) \\
\bold{S}(\bold{k}) & 0 \\
\end{pmatrix}
\begin{pmatrix}
\bold{b}_m \\
\bold{b}_m'\\
\end{pmatrix}
\ ,\nonumber\\
\eeqar
where $\bold{b}_m$ represents the baryon field with four different momenta,
\beqar
\bold{b}_m &=& \left(
		b_{m\bold{k}^{(1)}},\ 
		b_{m\bold{k}^{(2)}},\ 
		b_{m\bold{k}^{(3)}},\ 
		b_{m\bold{k}^{(4)}}
	\right)
\ ,
\nonumber\\
\bold{b}_m' &=& \bold{b}_{m+\beta/2}
\ .
\eeqar
The matrix $\bold{S}$ represents how these different momentum states
are connected through $\eta_j$,
\beq
\bold{S} =
-{i\over4}
\begin{pmatrix}
\sin k_1 & \sin k_2 & \sin k_3 & 0\\
\sin k_2 & -\sin k_1 & 0 & \sin k_3\\
\sin k_3 & 0 & -\sin k_1 & -\sin k_2\\
0 & \sin k_3 & -\sin k_2 & \sin k_1\\
\end{pmatrix}
\eeq
It is interesting to find that the square of $\bold{S}$ becomes
a $c$-number,
\begin{equation}
\bold{S}\cdot\bold{S}= -\frac{1}{16}s^2 \bold{1}
\ , \quad
(s^2 = \sum_{j=1}^{3}\sin^2k_j)
\ .
\end{equation}
Now we can evaluate the baryon integral,
\beqar
\exp(-\beta L^3 \FeffB)
&\equiv&
{\rm Det}V_B\int \Fint{b,\bar{b}}
	\exp\left[-\left(\bar{b},\VB^{-1} b\right)
		\right]
\nonumber\\
&=&
	{\rm Det}\left[
		1 +  \omega V_B
		\right]
\nonumber\\
&=&
\prod_{k_1=1}^{L/2}
\prod_{k_2=1}^{L/2}
\prod_{k_3=1}^{L}
\prod_{m=1}^{\beta/2}
{\rm det}
\Biggl[
\begin{pmatrix}
1 & \omega \bold{S} \\
\omega \bold{S} & 1 \\
\end{pmatrix}
\Biggr]
\nonumber\\
&=&
\prod_{k_1=1}^{L/2}
\prod_{k_2=1}^{L/2}
\prod_{k_3=1}^{L}
\prod_{m=1}^{\beta/2}
\left(
	1 + \omega^2 s^2/16
\right)^4
\nonumber\\
&=&
\prod_{\bold{k}}
\left(
	1 + \omega^2 s^2/16
\right)^{\beta/2}
\ .
\eeqar

For very large spatial lattice size $L$, 
we can replace the sum by the integral,
\beqar
&&\FeffB(\omega)
= - {1\over 2L^3} \sum_{\bold{k}} \log\left[1 + {\omega^2 s^2\over16}\right]
\nonumber\\
&&= - \frac{1}{2\pi^3} \int_{-\pi/2}^{\pi/2} dk_1 \int_{-\pi/2}^{\pi/2} dk_2 \int_{-\pi/2}^{\pi/2} dk_3
\log\left[1 + {\omega^2 s^2\over16}\right]
\nonumber\\
&&\simeq - \frac{a^{(b)}_0}{2}
\left({4\pi\over 3} \Lambda^3\right)^{-1}
\int_0^{\Lambda} 4\pi k^2 dk 
\log\left[1 + {\omega^2 k^2\over16} \right]
\nonumber\\
&&
= -a^{(b)}_0 f^{(b)}({\omega\Lambda\over4})
\ .
\eeqar
In the fourth line,
we have made an approximation
to replace the average in a box
to that in a sphere.
With this approximation, the effective free energy from 
baryon integral can be represented by a function, $f^{(b)}(x)$,
\beqar
&&f^{(b)}(x)\equiv
{3\over 2x^3}\int_0^x k^2 dk \log(1+k^2)
\nonumber\\
&=&\frac12 \log(1+x^2)
	- {1\over x^3}\left[
		\arctan{x} - x + {x^3\over 3}
	\right]
\ .
\eeqar
From numerical studies,
following normalization factor and the cut off,
\beq
a^{(b)}_0 = 1.0055\ ,\quad
\Lambda = {\pi\over 2} \times 1.01502\ ,
\eeq
are found to give a good global fit of $\FeffB$.

\comment{
Numerically, it is better to use a little modified function,
\beqar
\FeffB
 &\simeq& - f^{(b)}_1(c\Lambda)
\ ,\\
f^{(b)}_1(x) &=& \frac{a^{(b)}_1}{2} \log(1+x^2)
	- {a^{(b)}_2 \over x^3}\left[
		\arctan{x} - x + {x^3\over 3}
	\right]
\ ,\\
\Lambda &=& {\pi \over 2} \times 0.871192
\ ,\quad
a^{(b)}_1 = 1.0013\ ,\quad 
a^{(b)}_2 = 0.513501\ .
\eeqar
}

The baryon determinant $\Feff^{(b\mu)}$
in Eqs.~(\ref{Eq:Feff0b}) and (\ref{Eq:Feff0bv})
can be also evaluated by using a similar technique
shown here.
We show only the results here.
\beqar
&&
\Feff^{(b\mu)}(m;T,\mu)
\nonumber\\
&&\equiv -{T\over L^3}\log\mathrm{Det}\left(m/4 + V_{B\mu}\right)
\nonumber\\
&&= -{T\over 2 L^3}\sum_{k_0,\bold{k}}
	\log\left[
		m^2 + \sin^2(k_0 - 3i\mu) + s^2
		\right]
\nonumber\\
&&= -{T\over L^3}\sum_{\bold{k}}
	\log\left[
		\cosh(\beta\arcsinh\sqrt{s^2+m^2})
		+C_{3\mu}
		\right]
\ ,
\nonumber\\
\eeqar
where, $s^2 = \sum_{j=1}^3 \sin^2 k_j$.
In the numerical calculations in Subsec.~\ref{Subsec:Comp},
we have adopted the following approximation
assuming that the spatial lattice size $L$ is large enough
and that the average in a cubic box can be well approximated
by the average in a sphere,
\beqar
&&
\Feff^{(b\mu)}(m;T,\mu)
\nonumber\\
&&\simeq -a_0^{(b\mu)} T {3\over4\pi\Lambda^3}
	\int^\Lambda d\bold{k} \log\left[
			 C_b(\bold{k},m)
			+C_{3\mu}
			\right]
\ ,
\\
&&
C_b(\bold{k},m)
= \cosh\left[\beta\arcsinh\sqrt{\bold{k}^2+m^2}\right]
\ ,
\eeqar
where the simple ansatz $a_0^{(b\mu)}=1$ and $\Lambda=\pi/2$
are used in the numerical calculations shown in 
Subsec \ref{Subsec:Comp}.

\end{document}